\documentclass[acpd, manuscript]{copernicus_v}

\begin{document}

\title{OH level populations and accuracies of Einstein-A coefficients from
  hundreds of measured lines}


\Author[1,2]{Stefan}{Noll}
\Author[3]{Holger}{Winkler}
\Author[2]{Oleg}{Goussev}
\Author[4]{Bastian}{Proxauf}

\affil[1]{Institut für Physik, Universit\"at Augsburg, Augsburg, Germany}
\affil[2]{Deutsches Fernerkundungsdatenzentrum, Deutsches Zentrum f\"ur Luft-
  und Raumfahrt, We\ss{}ling-Oberpfaffenhofen, Germany}
\affil[3]{Institut f\"ur Umweltphysik, Universit\"at Bremen, Bremen, Germany}
\affil[4]{Max-Planck-Institut f\"ur Sonnensystemforschung, G\"ottingen,
  Germany}
  

\runningtitle{OH level populations and Einstein-A coefficients}

\runningauthor{S. Noll et al.}

\correspondence{S. Noll (stefan.noll@dlr.de)}

\received{}
\pubdiscuss{} 
\revised{}
\accepted{}
\published{}


\firstpage{1}

\maketitle

\begin{abstract}
OH airglow is an important nocturnal emission of the Earth's mesopause region.
As it is chemiluminescent radiation in a thin medium, the population
distribution over the various roto-vibrational OH energy levels of the
electronic ground state is not in local thermodynamic equilibrium (LTE). In
order to better understand these non-LTE effects, we studied hundreds of OH
lines in a high-quality mean spectrum based on observations with the
high-resolution Ultraviolet and Visual Echelle Spectrograph at Cerro Paranal
in Chile. Our derived populations cover vibrational levels between $v = 3$ and
9, rotational levels up to $N = 24$, and individual $\Lambda$-doublet
components when resolved. As the reliability of these results critically
depends on the Einstein-A coefficients used, we tested six different sets and
found clear systematic errors in all of them, especially for Q-branch lines
and individual $\Lambda$-doublet components. In order to minimise the
deviations in the populations for the same upper level, we used the most
promising coefficients from Brooke et al. (2016, JQSRT 168, 142) and further
improved them with an empirical correction approach. The resulting rotational
level populations show a clear bimodality for each $v$, which is characterised
by a probably fully thermalised cold component and a hot population where the
rotational temperature increases between $v = 9$ and 4 from about 700 to about
7,000\,K and the corresponding contribution to the total population at the
lowest $N$ decreases by an order of magnitude. The presence of the hot
populations causes non-LTE contributions to rotational temperatures at low
$N$, which can be estimated quite robustly based on the two-temperature model.
The bimodality is also clearly indicated by the dependence of the populations
on changes in the effective emission height of the OH emission layer. The
degree of thermalisation decreases with increasing layer height due to a
higher fraction of the hot component. Our high-quality population data are
promising with respect to a better understanding of the OH thermalisation
process. 
\end{abstract}


\introduction[Introduction]  
\label{sec:intro}

The nighttime emission of the Earth's atmosphere in the near-infrared is
dominated by hydroxyl (OH) airglow
\citep{meinel50,rousselot00,hanuschik03,noll12,noll15}, which originates in
the mesopause region in a layer with a width of about 8\,km and a typical peak
height of 87\,km \citep{baker88}. The various bright roto-vibrational bands of
the OH electronic ground state X$^2\Pi$ represent an important tracer for
atmospheric dynamics (especially wave propagation), ambient temperatures, and
chemical composition (especially atomic oxygen) at these high altitudes, which
are mostly probed by ground- and satellite-based remote sensing
\citep[e.g.][]{taylor97,beig03,savigny12,mlynczak13,reisin14,sedlak16,noll17}.
For these applications, it is crucial to understand the physical mechanisms
that lead to the observed line emission.

In the mesopause region, OH is mostly formed by the reaction of hydrogen and
ozone \citep{bates50,xu12}, which excites the electronic ground state up to
the ninth vibrational level $v$ \citep{charters71,llewellyn78,adler97}. The
nascent population distribution over the roto-vibrational levels is far from
local thermodynamic equilibrium (LTE). As the subsequent relaxation processes
by collisions with other atmospheric species are relatively slow compared to
the radiative lifetimes of the excited states
\citep[e.g.,][]{adler97,xu12,kalogerakis18,noll18b}, the OH emission bands
(which contribute to the vibrational relaxation) reveal strong non-LTE
effects. The vibrational level populations can be fitted as a function of
energy by an exponentially decreasing (i.e. Boltzmann-like)
distribution with a pseudo-temperature of around 10,000\,K
\citep{khomich08,noll15,hart19a}. Hence, OH bands with upper state vibrational
levels $v^{\prime}$ up to the highest nascent state can easily be measured.
Moreover, the rotational level populations for the different $v$ reveal high
overpopulations for high rotational states $N$ compared to the lowest three or
four levels under the assumption of a thermal distribution
\citep{pendleton89,pendleton93,dodd94,cosby07,oliva15,noll18b}.
The pseudo-temperatures for the high-$N$ populations achieve values up to
those found for the $v$ levels \citep{oliva15}. The theoretical explanation of
these populations especially for low $v$ is still uncertain
\citep{dodd94,kalogerakis18,noll18b} as their modelling suffers from
limitations in the data sets and uncertain input parameters (especially rate
coefficients for collisional transitions). It is usually assumed that the
ratios of lines related to the lowest $N$ of a fixed $v$ are sufficiently
close to LTE for a reliable estimate of the ambient temperature
\citep[e.g.][]{beig03}. However, this assumption appears to be insufficient
at least for the highest $v$, where deviations of several kelvins were found
\citep{noll16,noll18b}. In addition, small modifications in the set of
considered levels in terms of $N$ can already significantly change the
corresponding population temperature \citep{noll15}. 

A successful study of OH level populations requires accurate molecular
parameters, i.e. line wavelengths, level energies, and Einstein-A
coefficients. In particular, the latter suffer from relatively high
uncertainties despite numerous dedicated studies for their calculation
\citep[e.g.,][]{mies74,langhoff86,turnbull89,nelson90,goldman98,vanderloo07,
  brooke16} and evaluation
\citep[e.g.,][]{french00,pendleton02,cosby07,liu15,hart19b}. Apart from the
derivation of absolute OH level populations or densities
\citep{noll18b,hart19b}, the quality of these transition probabilities
especially affects OH-based temperature estimates
\citep{liu15,noll15,parihar17,hart19b} and abundance retrievals for species
like atomic oxygen \citep{mlynczak13,noll18b}. The persistent uncertainties
in the Einstein-A coefficients are obviously related to the molecular
structure of OH and the lack of adequate data for the calculation of the
molecular parameters
\citep{nelson90,pendleton02,cosby07,vanderloo07,brooke16}.

In order to improve our knowledge on OH level populations and Einstein-A
coefficients, high-quality measurements of a large number of OH lines and a
detailed analysis are required. We could perform such a study based on
high-resolution spectroscopic data taken with the Ultraviolet and Visual
Echelle Spectrograph \citep[UVES;][]{dekker00} at the Very Large Telescope at
Cerro Paranal in Chile (24.6$^{\circ}$\,S, 70.4$^{\circ}$\,W). A mean spectrum
of the highest-quality spectra (totalling 536 hours of exposure time) allowed
us to investigate 723 lines with upper vibrational levels $v^{\prime}$ between
3 and 9 in the optical and near-infrared regime in detail. In many cases, the
small $\Lambda$ doubling effect due to rotational--electronic perturbations
between the ground and excited electronic states \citep{pendleton02} was
resolved.

In Sect.~\ref{sec:data}, we describe the UVES data set. Then, we discuss the
data analysis involving the calculation of the mean spectrum, the measurement
of line intensities, and a check of the line positions
(Sect.~\ref{sec:analysis}). Section~\ref{sec:acoeff} discusses the
differences in the derived OH level populations for the Einstein-A
coefficients of \citet{mies74}, \citet{langhoff86}, \citet{turnbull89},
\citet{vanderloo08}, \citet{rothman13}, and \citet{brooke16}. Moreover, the
\citet{brooke16} reference is used as the basis for an empirical improvement
of the coefficients. The corresponding OH level populations are then
investigated in detail (Sect.~\ref{sec:ohpop}). This involves population
fitting, a study of the non-LTE contributions to rotational temperatures, and
the investigation of population differences caused by a change in the OH
emission altitude. Finally, we draw our conclusions
(Sect.~\ref{sec:conclusions}).

\section{Data set}\label{sec:data}

This study is based on so-called Phase~3 products of the astronomical echelle
spectrograph UVES \citep{dekker00} provided by the European Southern
Observatory. \citet{noll17} selected about 10,400 archived spectra taken
between April 2000 and March 2015, extracted the night-sky emission, and
performed a complex flux calibration procedure in order to investigate
long-term variations in the mesopause region based on OH emission. The studied
spectra comprise the wavelength range between 570 and 1040\,nm covered by two
set-ups centred on 760 and 860\,nm. Depending on the width of the entrance
slit, the spectral resolving power varied between 20,000 and 110,000. Hence,
these data are well suited for OH level population studies as they allow one
to measure numerous resolved emission lines.   

As the exposure time (between 1 and 125\,min) and the contamination of the
night-sky emission by the astronomical target (the slit length is only between
8 and 12$^{\prime\prime}$) are also strongly varying, it is important to focus on
spectra of sufficient quality, especially if very weak lines are studied.
The final sample of \citet{noll17}, who studied relatively bright P-branch
lines related to low rotational levels, included 3,113 suitable spectra. We
use an even smaller subsample of 2,299 spectra as the basis for this study. It
is related to the investigation of the faint K(D$_1$) potassium line at
769.9\,nm with a mean intensity of about 1\,R (rayleigh) by \citet{noll19}.
For that sample, the selected spectra were carefully checked around the
K(D$_1$) line. In order to be able to measure even fainter lines in the entire
wavelength regime and to have a homogeneous data set for the calculation of a
mean spectrum, we further reduced the sample. 45 spectra of the set-up centred
on 860\,nm were rejected as they showed severe flaws (wrong continuum levels)
below 730\,nm. This was not a problem for the potassium study. Moreover, we
increased the minimum exposure time from 10 to 45\,min and reduced the maximum
continuum limit around K(D$_1$) from 100 to 40\,R\,nm$^{-1}$. Finally, we only
considered spectra that were taken with the standard slit width of
1$^{\prime\prime}$, which corresponds to a resolving power of 42,000. 63\% of
the sample of \citet{noll19} was taken with this slit width.

The resulting sample consists of 533 high-quality spectra with a total
exposure time of 536 hours at a telescope with a diameter of the primary
mirror of 8\,m.

\section{Analysis}\label{sec:analysis}

\subsection{Mean spectrum}\label{sec:meanspec}

In order to calculate the probably best high-resolution airglow mean spectrum
in the covered wavelength regime so far, we first mapped the 533 selected
spectra (Sect.~\ref{sec:data}) with the flux calibration of \citet{noll17}
applied to a common wavelength grid from 560 to 1061\,nm with a step size of
1\,pm that well samples airglow lines, which have a full width at half maximum
of about 20\,pm close to 800\,nm. The mapping is necessary since each UVES
spectrum has its own wavelength grid. The set-up positioning appears to
have an uncertainty of the order of 1\,nm. Moreover, the original step sizes
vary from 1.8 to 5.2\,pm depending on the central wavelength of the set-up
(760 or 860\,nm), the chip (two chips with spectra separated by a small gap at
the central wavelength), and the pixel binning. Pixel pairs in dispersion
direction on the chips were merged for 85\% of the sample. The rest of the
data are unbinned. Before the mean calculation, the spectra were also scaled to
be representative of the zenith by using the van Rhijn correction
\citep{vanrhijn21} for a thin layer at an altitude of 90\,km. As the zenith
angles at the mid-exposure times vary from 3 to 64$^{\circ}$, this is a crucial
correction with factors between 0.46 and 1.00. These factors do not
significantly change across the entire OH emission layer, i.e. the choice of
the reference altitude is not critical.

\begin{figure*}[t]
\includegraphics[width=17cm]{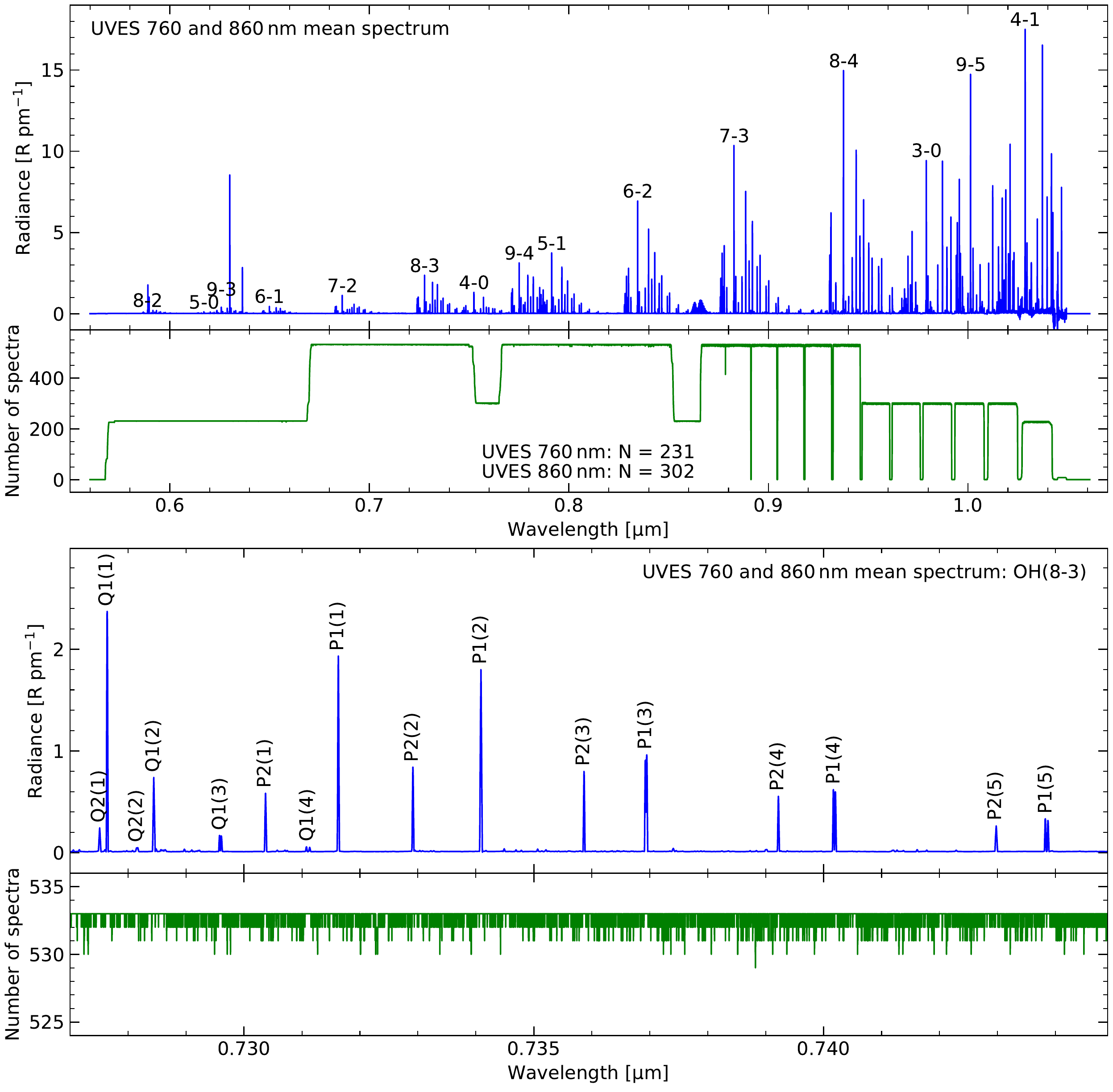}
\caption{UVES mean spectrum in rayleighs per picometre. The full spectral
  range with labelled OH bands is shown at the top. At the bottom, a narrower
  wavelength range focusing on the Q and P branch of \mbox{OH(8-3)} is
  plotted. The lines are labelled. The wavelength-dependent number of spectra
  involved in the calculation of the mean spectrum is indicated in additional
  subpanels. This number varies due to the use of two set-ups centred on 760
  and 860\,nm (with some variation) and the $\sigma$-clipping procedure for
  the mean calculation.}
\label{fig:spec}
\end{figure*}

The mean spectrum was calculated by means of a pixel-dependent
$\sigma$-clipping approach in order to avoid the contribution of strong
sporadic outliers due to technical issues or the contamination by an
astronomical target. As the threshold was set to 10 standard deviations,
statistical noise and natural variations of the airglow emission do not cause
the rejection of a spectrum at a certain pixel. The final number of considered
spectra at each wavelength after three iterations of the $\sigma$-clipping
is displayed in the lower panel of the upper plot in Fig.~\ref{fig:spec}. The
clipping only reduced the numbers by a few spectra. There is a trend towards
more rejections at longer wavelengths. The plot also reveals the impact of
the combination of the two set-ups centred on 760 and 860\,nm with 231 and 302
spectra, respectively. The gap between the spectra of the two chips of each
set-up and very narrow gaps between the spectral orders at long wavelengths
can also be seen. The rounded edges of the sample-related steps in the
histogram reflect the variation in the wavelength positioning of a certain
set-up. 

The mean spectrum in the upper plot in Fig.~\ref{fig:spec} shows 15 OH bands
marked by the upper and lower vibrational levels $v^{\prime}$ and
$v^{\prime\prime}$. Bands with $\Delta v = v^{\prime} - v^{\prime\prime}$
between 3 and 6 are covered. The band strength strongly increases from
\mbox{OH(5-0)} to \mbox{OH(4-1)} as bands with higher $v^{\prime}$ and lower
$\Delta v$ tend to be stronger in the covered wavelength regime. Each band is
split into the three branches R, Q, and P, which are characterised by changes
of the rotational quantum number $N$ of $-1$, 0, and $+1$. While the R and Q
branches at the short-wavelength side and central part of the band are
relatively compact, the P branch shows relatively wide spaces between the
lines. P-branch lines with high upper rotational quantum numbers $N^{\prime}$
are located at distinctly longer wavelengths than those with low $N^{\prime}$,
i.e. they are found in regions that are dominated by other OH bands. For this
reason, high-$N^{\prime}$ P-branch lines of the faint \mbox{OH(7-1)} band can
also be detected in the UVES mean spectrum.

The lower plot of Fig.~\ref{fig:spec} shows the narrow wavelength range
between 727 and 745\,nm to demonstrate the good spectral resolution. The
plotted range includes the full Q branch and the P branch up to lines with
$N^{\prime} = 5$ of \mbox{OH(8-3)}, a band of intermediate strength. The plot
clearly shows the splitting of each rotational state by spin--orbit coupling.
The Q$_1$ and P$_1$ lines ($F = 1$) related to the electronic substate
X$^2\Pi_{3/2}$ are well separated from the fainter Q$_2$ and P$_2$ lines
($F = 2$) of X$^2\Pi_{1/2}$. For the visible lines, the value of $F$ does not
change during the transition, i.e. $F^{\prime\prime} = F^{\prime}$. The
intercombination lines which show a change of $F$ are much fainter and are
therefore neglected in this study. The spectral resolving power of 42,000 is
sufficiently high for seeing $\Lambda$ doubling. The separation of both
components can already be found for Q$_1$ and P$_1$ lines with relatively low
$N^{\prime}$. In the lower plot of Fig.~\ref{fig:spec}, the largest separation
is visible for Q$_1$($N^{\prime} = 4$). It amounts to 55\,pm \citep{brooke16},
i.e. this $\Lambda$ doublet is fully resolved. Separations of more than
200\,pm are measurable for P$_1$ lines with $N^{\prime} \ge 11$. The faintest
marked $\Lambda$ doublets Q$_2$(2) and Q$_1$(4) have intensities between 1 and
2\,R. They can easily be measured in the UVES mean spectrum, which allows one
to also detect lines that are more than one order of magnitude fainter
(Sect.~\ref{sec:lineint}). For a general overview of lines (not only OH)
that can be accessed with UVES data, see the catalogue of \citet{cosby06}. It
is based on the night-sky atlas of \citet{hanuschik03}, which involves UVES
observations with a total exposure time of 9 hours in the red and
near-infrared wavelength range.

\subsection{Line intensities}\label{sec:lineint}

As it is the most comprehensive list of calculated OH lines so far, we used
the line wavelengths of \citet{brooke16} for the identification of lines in
the UVES mean spectrum (Sect.~\ref{sec:meanspec}) and the derivation of
their intensities. For this purpose, the calculated vacuum wavelengths were
converted into air wavelengths by means of the formula of \citet{edlen66} for
standard air, which works well for the UVES data. The default line integration
range was set to a width of about 2 resolution elements of the spectrograph
(Sect.~\ref{sec:data}), which is 40\,pm at 860\,nm, plus the separation of
the two $\Lambda$-doublet components. If the latter was wider than the 2
resolution elements, the components were measured independently. For an
optimal continuum subtraction, the two continuum points for a linear
interpolation across the line were defined manually as this approach can
better handle contaminations by nearby emissions and absorptions of other
lines than an automatic procedure, which was also tested. The wavelengths of
the selected continuum points were also used as limits of the integration
range. In particular, range modifications were necessary for lines at very
long wavelengths around 1\,$\mu$m, where the spectrograph causes extended line
wings. The resulting line intensities representing the zenith
(Sect.~\ref{sec:meanspec}) were also corrected for molecular absorption in
the lower atmosphere. The complex procedure involving high-resolution
radiative transfer calculations and water vapour measurements in the
astronomical target spectra is described by \citet{noll17}. Further details
are given by \citet{noll15}. For the correction of the measured intensities,
the derived line transmission values for the 533 individual spectra were
averaged. The resulting mean absorption of the measured $\Lambda$ doublets was
3\% and only 6\% of the doublets was attenuated by more than 10\%. Hence, the
related intensity uncertainty after the correction, which can reduce the
absorption by up to an order of magnitude, is negligible for most lines.

As illustrated in Fig.~\ref{fig:spec}, the mean spectrum is composed of
UVES spectra of two different set-ups centred on 760 and 860\,nm. The
wavelength shift between both set-ups causes changes in the data properties
depending on wavelength. In order to minimise the impact of these changes on
the measured line intensities, we investigated and corrected two effects:
long-term variations in the OH line intensity and flux calibration errors.
The former are important since the two UVES set-ups cover very different
parts of the sample-related period from May 2000 to July 2014. Before December
2004, there were only observations with the 860\,nm set-up \citep{noll17}. On
the other hand, spectra of this set-up are only present in the selected sample
until May 2010. This results in mean 10.7\,cm solar radio fluxes
\citep{tapping13} for an averaging period of 27 days of 102 solar flux units
(sfu) for the 760\,nm set-up and 140\,sfu for the 860\,nm set-up. According to
\citet{noll17} (also based on UVES data), the mean solar cycle effect for
$v^{\prime}$ between 5 and 9 is $16.1 \pm 1.9$\,\% per 100\,sfu. There is no
significant change with $v^{\prime}$. We took this mean percentage, the
set-up-specific mean solar radio fluxes, and the wavelength-dependent fraction
of 760 and 860\,nm spectra to correct the OH line intensities to be
representative of the mean solar radio flux of the full sample of 123\,sfu.
The intensity corrections were up to a few per cent with line-dependent
differences characterised by a standard deviation of 2.2\%. We did not
consider the impact of a possible linear long-term trend as it is not
significant for the UVES data \citep{noll17}. In order to test the flux
calibration, we calculated mean spectra for each set-up and also measured line
intensities. The latter was performed automatically by using the same
wavelengths for the line integration as in the case of the mean spectrum of
the full sample. The continuum was measured in narrow intervals (about 0.25
resolution elements, i.e. 5\,pm wide at 860\,nm) around these limiting
positions. The wavelength-dependent intensity ratios for the two set-ups were
then used to derive correction factors depending on set-up and chip. Taking
the wavelength range between the set-up gaps around 760 and 860\,nm as the
reference, we found correction factors close to 1 with a standard deviation of
2.7\%, which is consistent with the relative flux calibration uncertainty of
about 2\% for the UVES data set reported by \citet{noll17}. Combining the
solar activity and flux calibration correction, the resulting standard
deviation is only 1.7\% as both effects partly cancel out.

The quality of the final line intensities was indicated by a flag consisting
of a primary and a secondary classifier. Each classifier is represented by a
digit between 0 and 3. A value of 3 corresponds to a reliable measurement of
the entire $\Lambda$ doublet, which requires symmetric line emission and a
featureless underlying continuum, 1 and 2 refer to reliable measurements only
for the $\Lambda$-doublet component with e or f parity in the upper state, and
0 marks uncertainties for both components. For unresolved doublets, only 0 and
3 are possible digits. The introduction of the secondary classifier allows for
a finer classification scheme. For example, the combined classes 30 and 03 can
be used for ambiguous cases. Reasons for measurement uncertainties are
obvious or possible blends with other emission lines, regions of significant
absorptions in the continuum (often combined with very low transmission at the
position of the line), and insufficient signal-to-noise ratio in the case of
very weak lines.

\begin{figure}[t]
\includegraphics[width=8.3cm]{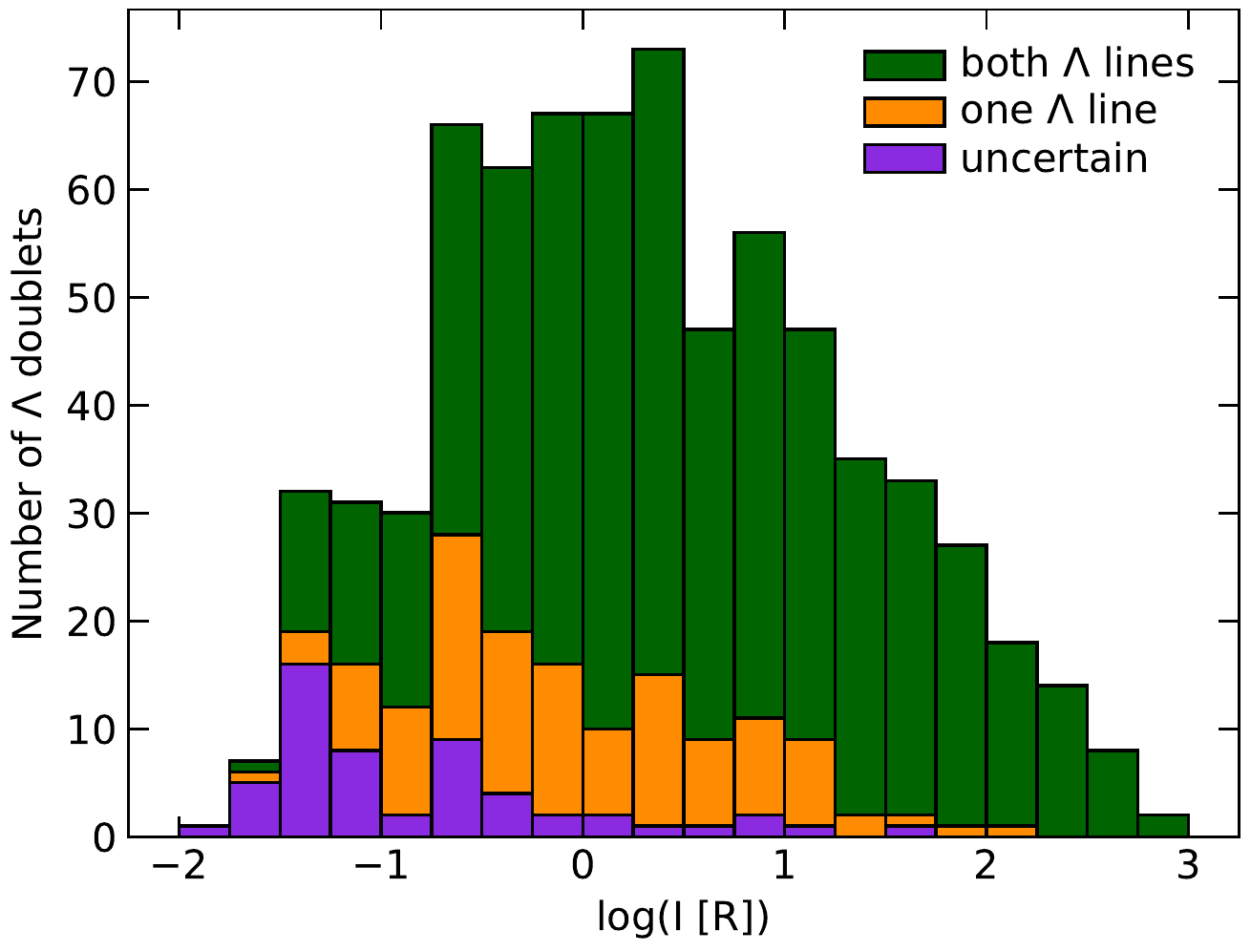}
\caption{Distribution of the decadal logarithm of the measured OH
  $\Lambda$-doublet intensities in rayleighs. Three categories are indicated:
  reliable measurement of both $\Lambda$-doublet components (detached or
  unresolved) in green (class 3), reliable measurement of only one component
  in orange (classes 1 and 2), and uncertain intensities in violet (class 0
  without 00).}
\label{fig:I_histo}
\end{figure}

Figure~\ref{fig:I_histo} shows a histogram of the measured lines depending on
the decadal logarithm of the intensity in rayleighs. In total, 723 $\Lambda$
doublets are included. This neglects 13 measurements with digit 0 for the
primary and secondary classifier. These lines were not further used in this
study. Other potential lines could not be measured due to a blend with a
stronger line, no detection, or a line wavelength within the order gaps in
the near-infrared (Fig.~\ref{fig:spec}). The measured intensities of the 723
$\Lambda$ doublets range from 0.01 to 600\,R, i.e. they comprise almost 5
orders of magnitude. Intensities around 1\,R are most abundant. The median is
1.7\,R. There is a conspicuous drop in the occurrence frequency below about
0.2\,R, which suggests a strongly increasing incompleteness of detections for
fainter lines. The intensity of weaker $\Lambda$ doublets also tends to be
more uncertain as the intensity distributions for the different quality
classes show. For the classes 3, 1+2, and 0, the median intensities are 2.6,
0.83, and 0.086\,R. 546 doublets or 76\% belong to class 3, where the two
components were measured independently in 34\% of the cases. There are 122
doublets (17\%) with only one reliable component (1+2). Finally, there are 55
cases (8\%) with class 0, 75\% of them with resolved doublets. Considering
that detached components require independent line measurements (350 cases),
the total number of measurements for the data in Figure~\ref{fig:I_histo}
amounts to 1,073. 

\begin{figure}[t]
\includegraphics[width=8.3cm]{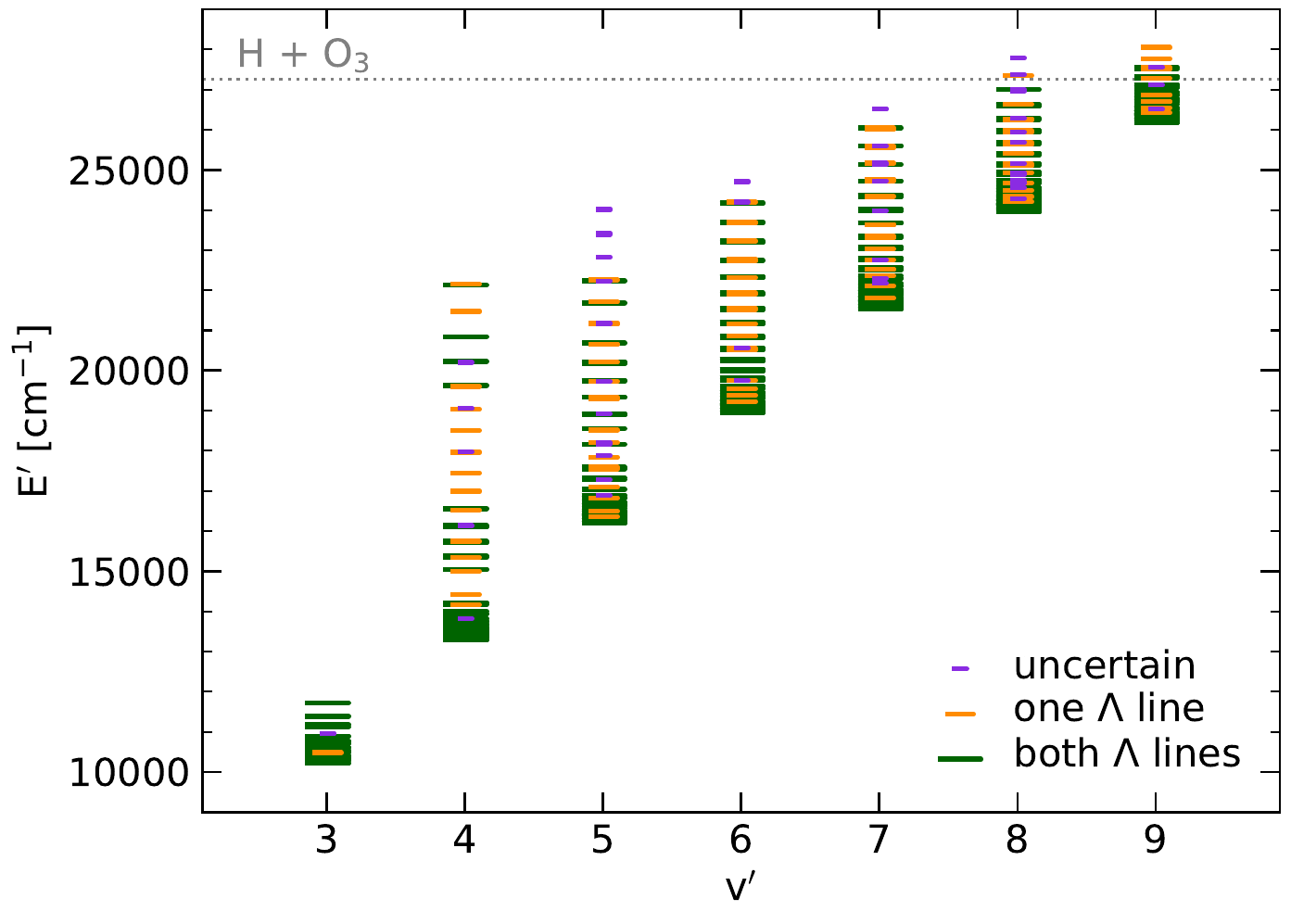}
\caption{Distribution of the upper level energies $E^{\prime}$ of the measured
  roto-vibrational transitions in inverse centimetres separated for each upper
  vibrational level $v^{\prime}$. The same categories as in
  Fig.~\ref{fig:I_histo} are marked by different colours and line lengths (see
  legend). The likely exothermicity limit of the hydrogen--ozone reaction is
  marked by the grey dotted line.}
\label{fig:Ep_vp}
\end{figure}

The 723 studied $\Lambda$ doublets probe 236 different upper states
characterised by $v^{\prime}$, $N^{\prime}$, and $F^{\prime}$. Up to nine doublets
contribute to the population data for a certain level. The distribution of
level energies $E^{\prime}$ depending on $v^{\prime}$ is shown in
Fig.~\ref{fig:Ep_vp}. The energies range from 10,211 to 28,051\,cm$^{-1}$.
Except for $v^{\prime} = 3$, where $N^{\prime}$ only up to 9 could be measured
(mainly due to the wavelength limitations of the UVES data), wide ranges of
$E^{\prime}$ are covered by the data for the different $v^{\prime}$. A maximum
energy range of 8,861\,cm$^{-1}$ is achieved for $v^{\prime} = 4$. This is
possible due to $N^{\prime}$ up to 24. The energy ranges shrink for higher
$v^{\prime}$ due to a steeper decrease of the line intensities with increasing
$N^{\prime}$, which reduces the detectability of high-$N^{\prime}$ lines. An
important reason for this is certainly the closer exothermicity limit of the
hydrogen--ozone reaction, which produces the excited
OH. Nevertheless, there are nine levels above this limit if we assume 3.38\,eV
\citep{cosby07}, i.e. about 27,260\,cm$^{-1}$ \citep{noll18b}. This suggests
that the kinetic energy involved in the reaction is also important to
populate the OH roto-vibrational levels. For the excitation of the highest
level found ($v^{\prime} = 9$, $N^{\prime} = 12$, $F^{\prime} = 1$), about
800\,cm$^{-1}$ of additional energy would be needed. Figure~\ref{fig:Ep_vp}
also provides the primary quality classes for the lines related to the
displayed states. 79\% of the levels are covered by at least one line with
class 3. An exclusive class 0 contribution is found for only 16 states.
However, excluding uncertain lines can reduce the $E^{\prime}$ range for a given
$v^{\prime}$. In particular, the maximum $N^{\prime}$ for $v^{\prime} = 5$ shows a
decrease from 23 to 20 in this case.

\subsection{Line positions}\label{sec:linepos}

The high resolving power of 42,000 of the UVES data used allows for a check of
the quality of the input line positions, which were taken from
\citet{brooke16} and were converted to standard air using the formula of
\citet{edlen66}. The default positioning of the integration ranges for the
line intensity measurements described in Sect.~\ref{sec:lineint} worked well
in most cases. However, significant shifts were necessary for some
high-$N^{\prime}$ lines. For a systematic study of these offsets, we took the
manually adapted integration windows to calculate the intensity-weighted
centroid wavelength for each line.

\begin{figure}[t]
\includegraphics[width=8.3cm]{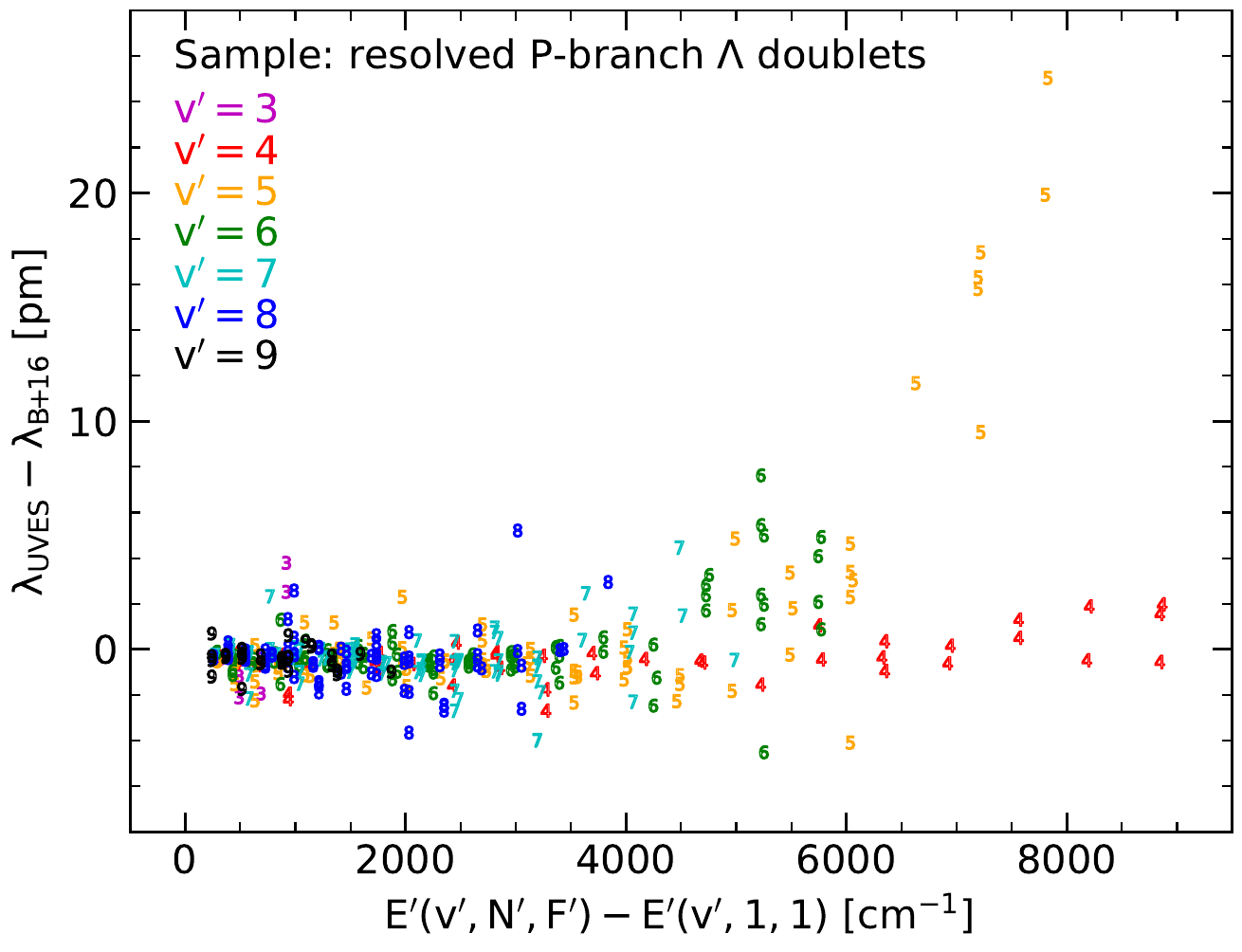}
\caption{Wavelength difference between the UVES-based measurement of isolated
  P-branch lines (resolved $\Lambda$ doublets) and the corresponding position
  provided by \citet{brooke16} (similar to HITRAN) in picometres as a function
  of the energy of the upper state of the transition $E^{\prime}$ with respect
  to the lowest energy of the corresponding vibrational level $v^{\prime}$ in
  inverse centimetres. The coloured markers indicate $v^{\prime}$.}
\label{fig:dlam_dEp}
\end{figure}

In Fig.~\ref{fig:dlam_dEp}, we show the difference between observed and model
wavelengths in picometres as a function of upper state energy (neglecting the
vibrational energy) for 406 individually measured $\Lambda$-doublet components
of the P branch. This selection rejects unresolved or only partly resolved
$\Lambda$ doublets and lines with uncertain central wavelengths due to
blending with other lines. The remaining 66 Q-branch and 67 R-branch
$\Lambda$-doublet components are not plotted as they only probe
$\Delta E^{\prime}$ up to 5,000\,cm$^{-1}$ and are essentially consistent with
the P-branch lines, which tend to have higher signal-to-noise ratios. The plot
shows for all $v^{\prime}$ a very good agreement of observed and modelled
wavelengths in the case of low energies. For $\Delta E^{\prime}$ lower than
2,000\,cm$^{-1}$, the mean value and standard deviation are $-0.4$ and
$0.7$\,pm, respectively. The systematic offset is much less than the original
pixel size in the UVES spectra (Sect.~\ref{sec:meanspec}). Hence, it can be
caused by uncertainties in the wavelength calibration. Moreover, the
assumption of standard air conditions (1013\,hPa, 288\,K, and no H$_2$O) for
the UVES instrument might cause a part of the offset. Beyond 3,000\,cm$^{-1}$,
the displayed wavelength offsets show an increasing scatter. In part, this is
caused by the higher measurement uncertainties for the fainter lines, but
there are also clear trends depending on $v^{\prime}$. In general, the
difference between the measured and theoretical line wavelengths increases
with $\Delta E^{\prime}$. This increase appears to be stronger for higher
$v^{\prime}$. While the change for $v^{\prime} = 4$ is only about 1\,pm at
around 8,000\,cm$^{-1}$, it is about 20\,pm for $v^{\prime} = 5$. For higher
$v^{\prime}$ (at least for 6 and 7), the increase of the offsets appears to be
even stronger. However, as the covered energy range decreases as well, the
maximum offsets only amount to a few picometres. Hence, only the measured
shifts for $v^{\prime} = 5$ and $N^{\prime}$ of 22 and 23 are of the order of a
spectral resolution element. For all other detected lines, the quality of the
theoretical line positions is much better. 

The discussed results are for the line wavelengths published by
\citet{brooke16}. As the HITRAN line database \citep{gordon17} is more
frequently used, we also calculated the wavelength shifts for those data. We
took the version HITRAN2012 \citep{rothman13}, which does not differ from the
more recent version HITRAN2016 \citep{gordon17} in terms of the OH data. The
results are very similar to those in Fig.~\ref{fig:dlam_dEp}. Strong
deviations above 10\,pm are found for the same small sample of lines, although
\mbox{OH(5-1)P$_1$(23)} is missing in the HITRAN database. The mean difference
between the line wavelengths from \citet{rothman13} and \citet{brooke16} is
0.07\,pm. The standard deviation only amounts to 0.40\,pm. 

Based on the UVES mean spectrum of \citet{hanuschik03}
(Sect.~\ref{sec:meanspec}), the accuracy of OH line wavelengths was already
investigated by \citet{cosby06}. Their theoretical line positions originate
from \citet{cosby00} but should be similar to \citet{goldman98}, the basis of
the OH data in HITRAN, for low rotational levels. For higher rotational
levels, the line wavelengths calculated by \citet{cosby00} should be more
precise. Indeed, although the spectrum of \citet{hanuschik03} is noisier, the
critical \mbox{OH(5-1)} lines do not show clear systematic offsets. However,
all OH data indicate a mean shift of the UVES-based wavelengths of about
$+1.0$\,pm. This offset is similar to those for atomic and molecular oxygen in
the same wavelength range, which were also measured by \citet{cosby06}. Thus,
systematic errors in the wavelength calibration are the most likely
explanation. In comparison, our measurements result in a negative mean offset.
This could be caused by differences in the UVES sample, data processing, and
analysis.

\section{Einstein-A coefficients}\label{sec:acoeff}

\subsection{Full OH level populations}\label{sec:fullpop}

The intensities $I_{i^{\prime}i^{\prime\prime}}$ derived in Sect.~\ref{sec:lineint},
where $i^{\prime}$ and $i^{\prime\prime}$ are the upper and lower states of the
roto-vibrational transition, can be converted into level populations by
dividing Einstein coefficients $A_{i^{\prime}i^{\prime\prime}}$. For a visualisation,
these populations are usually normalised by dividing the statistical weight
(i.e. the degeneracy) of the upper state $g^{\prime} = g_{i^{\prime}}$ and then
logarithmised. Following
\citet{noll15}, we define
\begin{equation}\label{eq:ydef}
  y :=
  \ln\bigg(\frac{I_{i^{\prime}i^{\prime\prime}}}{A_{i^{\prime}i^{\prime\prime}}
    g^{\prime}}\bigg),
\end{equation}
where the intensity is given in rayleighs and the Einstein-A coefficients are
provided in inverse seconds, which is consistent with population column
densities in units of $10^{6}$\,cm$^{-2}$ \citep{noll18b}. For $\Lambda$
doublets, $i$ is characterised by the vibrational level $v$, rotational level
$N$, and electronic substate $F$. In the case of individual components, the
parity $p$ (i.e. e or f) is also a parameter and $g^{\prime}$ is half as large
as for the doublet.

\begin{figure*}[t]
\includegraphics[width=17cm]{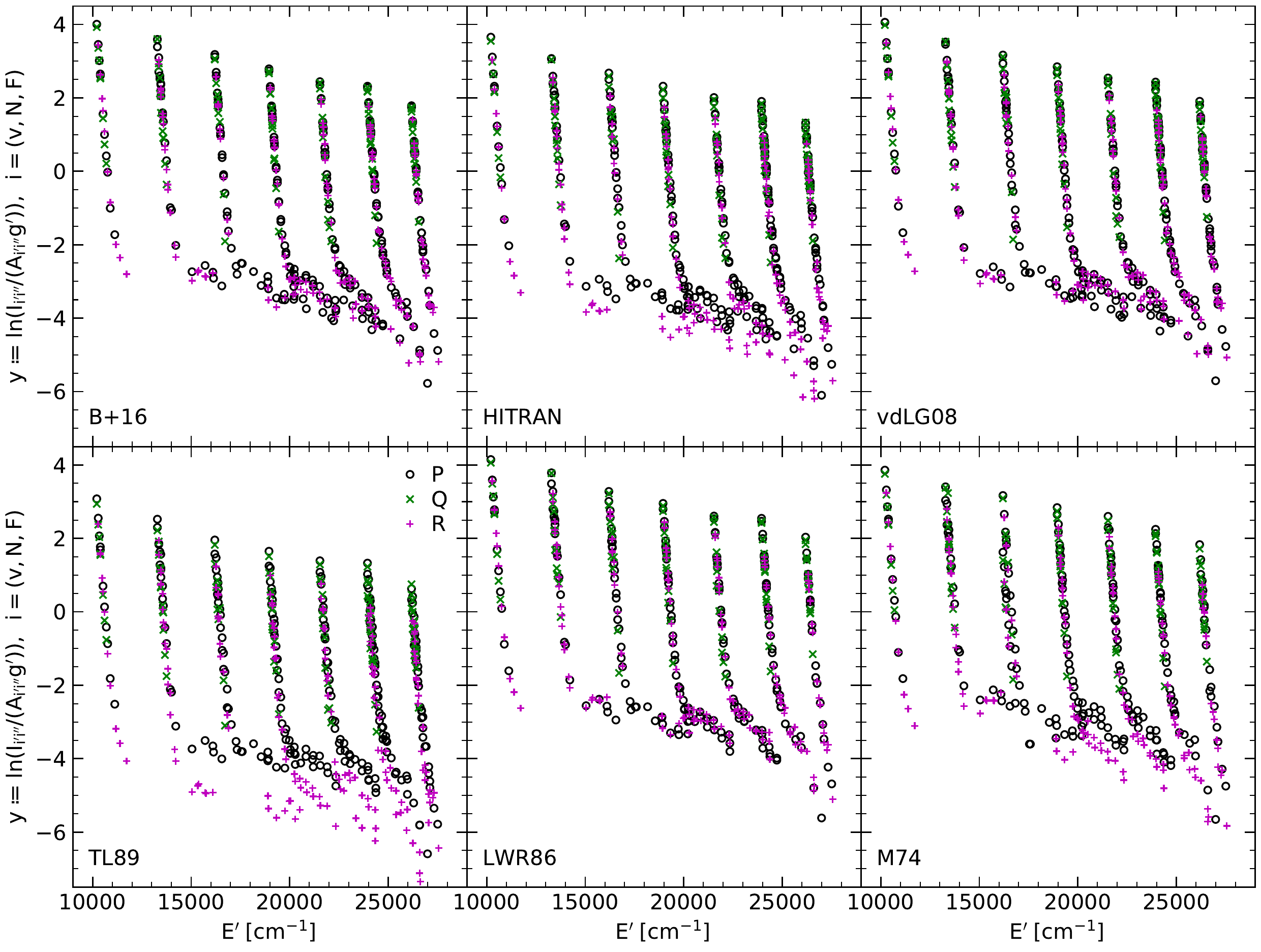}
\caption{Distribution of logarithmic OH level populations $y$, i.e. line
  intensity in rayleighs divided by Einstein-A coefficient in inverse seconds
  and the statistical weight of the upper level $g^{\prime}$, as a function
  of the upper level energy $E^{\prime}$ in inverse centimetres. The six panels
  show the results for the reliable $\Lambda$-doublet measurements (see
  Fig.~\ref{fig:I_histo}) neglecting \mbox{OH(7-1)} (two lines) for different
  sets of Einstein-A coefficients: \citet{brooke16} (\mbox{B+16}), HITRAN
  \citep{rothman13}, \citet{vanderloo08} (vdLG08), \citet{turnbull89} (TL89),
  \citet{langhoff86} (LWR86), and \citet{mies74} (M74). The number of plotted
  data points is smaller for older sets due to their limitations in the
  coverage of faint bands and high rotational levels. The populations derived
  from different branches (P, Q, and R) are highlighted by different symbols
  and colours (see legend in lower left panel).}
\label{fig:y_Ep}
\end{figure*}

Apart from the uncertainties in the line intensities, the quality of the
resulting populations also depends on the reliability of the Einstein-A
coefficients. As already mentioned in Sect.~\ref{sec:intro}, the latter is
not satisfactory as the available sets differ quite significantly. With our
large sample of energy levels, where the population of each state can be
derived from up to nine different lines, we can carry out a comprehensive
comparison of Einstein-A coefficients. As a reference set, we take the
coefficients calculated by \citet{brooke16} (\mbox{B+16}), who provide the
most recent and largest set of OH line parameters. The left upper panel of
Fig.~\ref{fig:y_Ep} shows the corresponding $y$ for the 544 reliable
$\Lambda$ doublets of class 3 (neglecting \mbox{OH(7-1)}, i.e. two doublets)
as a function of the energy of the upper state $E^{\prime}$. The distribution
of populations displays the well-known pattern of steep population decreases
for low $N^{\prime}$ and weaker population gradients for higher $N^{\prime}$
\citep{pendleton89,pendleton93,cosby07,oliva15,kalogerakis18,noll18b}.
Moreover, it indicates the expected decrease of populations for higher
$v^{\prime}$ with a remarkable exception for $v^{\prime} = 8$
\citep{cosby07,noll15}. The latter is a signature of the nascent OH level
population distribution, which mainly occupies $v^{\prime}  = 8$ and 9. The
population properties will be discussed in more detail in
Sect.~\ref{sec:ohpop}.

It is now important to know how robust the observed pattern is with respect
to changes in the set of Einstein-A coefficients. For this purpose, we also
consider the HITRAN database with the version from 2012 \citep{rothman13} (see
also Sect.~\ref{sec:linepos}), which is mainly based on the calculations of
\citet{goldman98} for OH. Moreover, we use the coefficients from
\citet{vanderloo08} (vdLG08), i.e. the corrected version of
\citet{vanderloo07}, \citet{turnbull89} (TL89), \citet{langhoff86} (LWR86),
and \citet{mies74} (M74). We neglect the also still popular data from
\citet{nelson90} as their line list only focuses on low $\Delta v$ and low
$N^{\prime}$. Except for \mbox{B+16}, our selection of sets agrees with
\citet{liu15} and \citet{hart19b}, who studied the impact of Einstein-A
coefficients on the populations of low rotational levels. Other comparisons
used a smaller number of sets
\citep{french00,cosby07,noll15,parihar17,noll18b}. Note that the three oldest
sets lack a significant number of the measured 723 $\Lambda$ doublets. The set
of \citet{turnbull89} only includes 634 doublets with maximum $N^{\prime}$
between 13 (R$_1$ branch) and 15 (Q$_2$ and P$_2$ branches). In the case of
LWR86 and M74, the $N^{\prime}$-related limits are higher by 1 but the bands
with $\Delta v = 6$, i.e. essentially \mbox{OH(8-2)} and \mbox{OH(9-3)}, are
not covered. The number of doublets is therefore only 566 or 78\% of the full
sample.

\begin{table*}[t]
\caption{Comparison of populations $y$ and population ratios $\Delta y$ with
  respect to changes in branch and $v^{\prime\prime}$ for six sets of
  Einstein-A coefficients}
\begin{tabular}{lccccccc}
\tophline
Set & $\langle y \rangle^\mathrm{a}$ &
$\langle \Delta y \rangle^\mathrm{b}$ &
$\langle \Delta y \rangle$ &
$\langle |\Delta y| \rangle^\mathrm{c}$ &
$\langle |\Delta y| \rangle$ &
$\langle |\Delta y| \rangle$ &
$\langle |\Delta y| \rangle$ \\
& & $\Delta N^\mathrm{\,d}$ & $\Delta N$ & $v^{\prime\prime\,e}$ &
$v^{\prime\prime}$ & $v^{\prime\prime}$ & $v^{\prime\prime}$ \\
& & Q$-$P & R$-$P & & P & Q & R \\
\middlehline
\mbox{B+16} & $-0.18$ & $-0.25$ & $-0.08$ & 0.11 & 0.08 & 0.17 & 0.10 \\
HITRAN & $-0.69$ & $-0.31$ & $-0.28$ & 0.19 & 0.12 & 0.27 & 0.27 \\
vdLG08 & $-0.13$ & $-0.25$ & $-0.06$ & 0.17 & 0.13 & 0.23 & 0.19 \\
TL89 & $-1.43$ & $-0.37$ & $-0.51$ & 0.38 & 0.28 & 0.44 & 0.53 \\
LWR86 & $-0.02$ & $-0.25$ & $-0.02$ & 0.18 & 0.14 & 0.25 & 0.20 \\
M74 & $-0.33$ & $-0.27$ & $-0.24$ & 0.41 & 0.42 & 0.36 & 0.43 \\
$N_\mathrm{sel}$$^\mathrm{\,f}$ & 416 & 82 & 96 & 127 & 65 & 30 & 32 \\
\bottomhline
\end{tabular}
\belowtable{
\begin{list}{}{}
\item[$^\mathrm{a}$] mean logarithmic level population $y$
\item[$^\mathrm{b}$] mean difference of logarithmic level populations
\item[$^\mathrm{c}$] mean absolute difference of logarithmic level populations
\item[$^\mathrm{d}$] change in branch (Q$-$P and R$-$P)
\item[$^\mathrm{e}$] change in $v^{\prime\prime}$ for all and individual branches
  (P, Q, R)
\item[$^\mathrm{f}$] number of selected $\Lambda$ doublets (present in all sets
  of Einstein-A coefficients)
\end{list}
}
\label{tab:compAset}
\end{table*}

Figure~\ref{fig:y_Ep} reveals clear discrepancies between the populations for
the six investigated sets of Einstein-A coefficients. The general structure of
the distribution is similar but the $y$ values are shifted. Taking 416
$\Lambda$ doublets with $N^{\prime} \le 12$ and $\Delta v \le 5$, which are
present in all six sets, we find mean $y$ between $-1.43$ for TL89 and $-0.02$
for LWR86 (Table~\ref{tab:compAset}). This corresponds to an unsatisfactorily
large population ratio of about 4.1. Substituting the extreme TL89 $y$ value
by the next highest one of $-0.69$ for HITRAN, the ratio is still about 1.9.
The coefficients of M74, \mbox{B+16}, and vdLG08 result in intermediate mean
$y$ values of $-0.33$, $-0.18$, and $-0.13$, respectively.

Any estimate of absolute OH level populations by means of OH line intensities
will be highly uncertain with these results if the quality of the Einstein-A
coefficients used cannot be evaluated. Tests of the accuracy of the measured
absolute populations require an alternative calculation which is less
sensitive to the choice of the set of molecular parameters. Using a kinetic
model for chemical OH production and excitation relaxation via collisions and
radiative transitions is a solution, although various required rate
coefficients and molecular abundances (especially for atomic oxygen) are quite
uncertain. \citet{noll18b} found that the higher populations related to
\mbox{B+16} tend to be more reliable than the lower ones based on HITRAN; this
was for $v = 9$ based on OH line intensities from UVES and pressure,
temperature, and molecular abundance profiles from the satellite-based
Sounding of the Atmosphere using Broadband Emission Radiometry (SABER)
instrument \citep{russell99} and the empirical atmospheric NRLMISE-00 model
\citep{picone02}. Hence, at least the very low populations related to TL89
appear to be quite unlikely.

\subsection{Detailed population comparisons}\label{sec:comparisons}

\begin{figure*}[t]
\includegraphics[width=17cm]{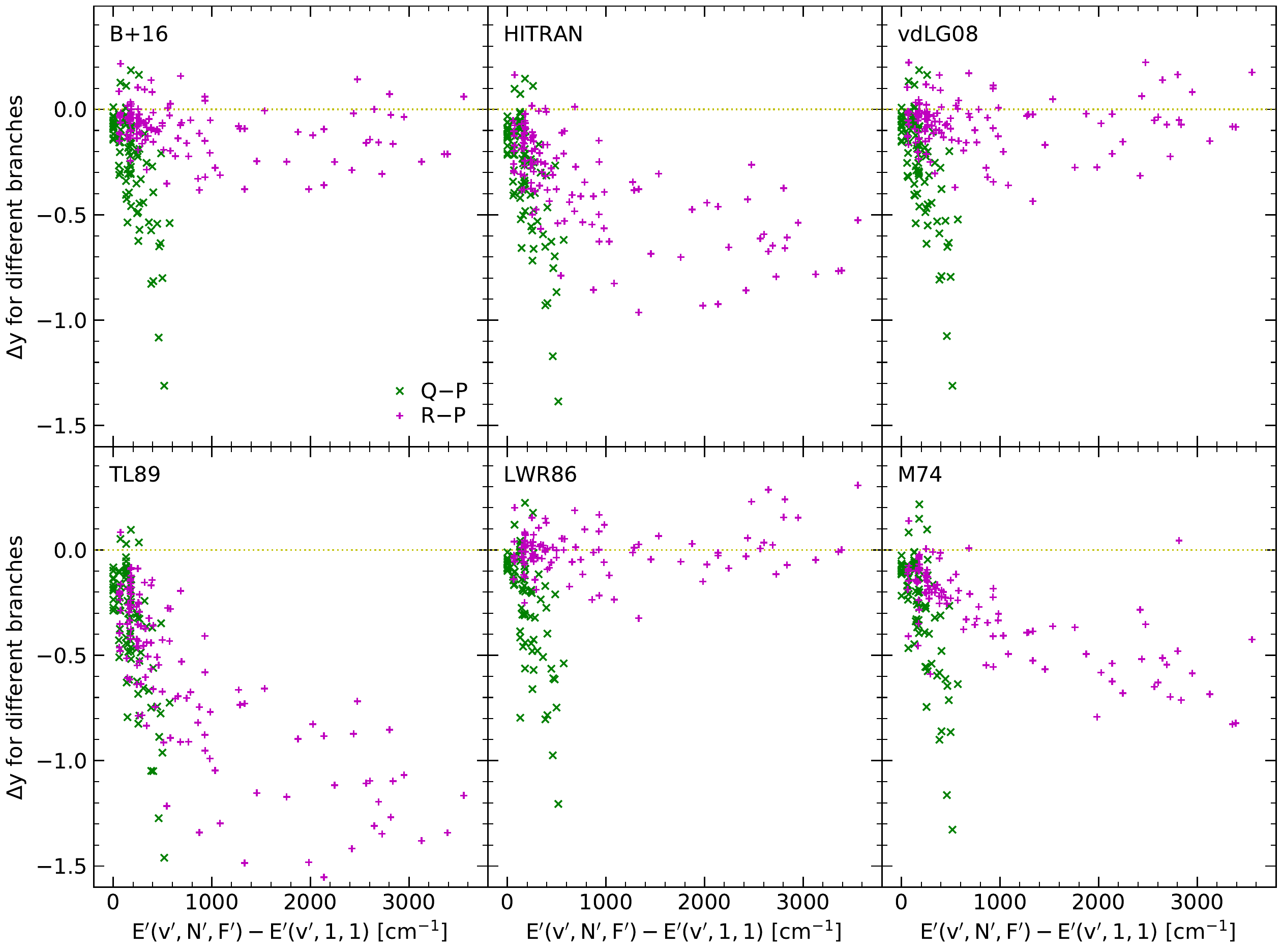}
\caption{Difference in logarithmic OH level populations $\Delta y$ by the
  change of the branch of the measured transitions from P to Q or R as a
  function of the upper state energy $E^{\prime}$ relative to the
  $v^{\prime}$-specific zero point (corresponding to $N^{\prime} = 1$ and
  $F^{\prime} = 1$) in inverse centimetres for six sets of Einstein-A
  coefficients. For more details, see legend and Fig.~\ref{fig:y_Ep}.}
\label{fig:dy_dEp_branch}
\end{figure*}

Figure~\ref{fig:y_Ep} marks the populations derived from lines of the R, Q,
and P branches by different symbols. A good set of Einstein-A coefficients
should result in similar population distributions for the three branches.
However, the TL89 $y$ values for the R branch are distinctly below those of
the P branch. A similar but weaker effect can also be seen for the data
related to HITRAN and M74. In order to study these discrepancies in more
detail, we calculated differences $\Delta y$ for lines with the same upper
state but different branches. We focus on Q versus P branch and R versus P
branch. The corresponding results for the six sets of Einstein-A coefficients
and reliable $\Lambda$ doublets of class 3 are shown as a function of
$E^{\prime}$ relative to the lowest energy for a given $v^{\prime}$ in
Fig.~\ref{fig:dy_dEp_branch}. For \mbox{B+16} Einstein-A coefficients, 219
population ratios $\Delta y$ are plotted.

All sets indicate unsatisfactory ratios for the comparison of Q- and P-related
populations, especially for high $\Delta E^{\prime}$ or $N^{\prime}$, where
$\Delta y$ can be lower than $-1$. The mean $\Delta y$ for a subsample with
$N^{\prime} \le 12$ and $\Delta v \le 5$ (c.f. Sect.~\ref{sec:fullpop}) are
between $-0.37$ for TL89 and $-0.25$ for \mbox{B+16}, vdLG08, and LWR86
(Table~\ref{tab:compAset}), i.e. the different sets fail in a similar way.
According to the theoretical considerations of \citet{pendleton02} triggered
by the \mbox{OH(6-2)} line intensity ratios measured by \citet{french00}, this
can be explained by the general negligence of orbital angular momentum
uncoupling, which is related to rotational--electronic mixing of the
electronic ground state X$^2\Pi$ and the first excited state A$^2\Sigma^+$,
for the calculation of the available Einstein-A coefficients. OH line
measurements in near-infrared spectroscopic data from the Nordic Optical
Telescope at La Palma (Spain) by \citet{franzen19} indicate that too low
Q-branch populations or too high Einstein-A coefficients (based on HITRAN) are
also an issue for OH bands with $\Delta v = 2$ and 3 not covered by our study.

The comparison of populations based on R- and P-branch lines reveals a more
complex situation than for the Q-branch data. All sets of Einstein-A
coefficients show negative mean $\Delta y$, i.e. lower R-related populations
on average (Table~\ref{tab:compAset}). However, the range is relatively wide
with values between $-0.51$ for TL89 and $-0.02$ for LWR86. The latter is the
only satisfactory set for this comparison at low $\Delta E^{\prime}$ as it was
already found by \citet{french00} for \mbox{OH(6-2)} low-$N^{\prime}$ lines.
For levels with high rotational energy, $\Delta y$ tends to be positive. For
the other sets, Fig.~\ref{fig:dy_dEp_branch} shows a clear decrease of
$\Delta y$ with increasing $\Delta E^{\prime}$ at least below 1,000\,cm$^{-1}$.
For example, the most recent set \mbox{B+16} shows mean $\Delta y$ of $-0.04$
and $-0.14$ below and above 400\,cm$^{-1}$, respectively. At high
$\Delta E^{\prime}$, the negative trend appears to vanish for \mbox{B+16},
HITRAN, and vdLG08, the latter even indicating an increase at the highest
energies as in the case of LWR86. The especially bad performance of the
TL89, HITRAN, and M74 coefficients is probably related to an underestimation
of the vibration--rotation interaction \citep{pendleton02}.

\begin{figure}[t]
\includegraphics[width=8.3cm]{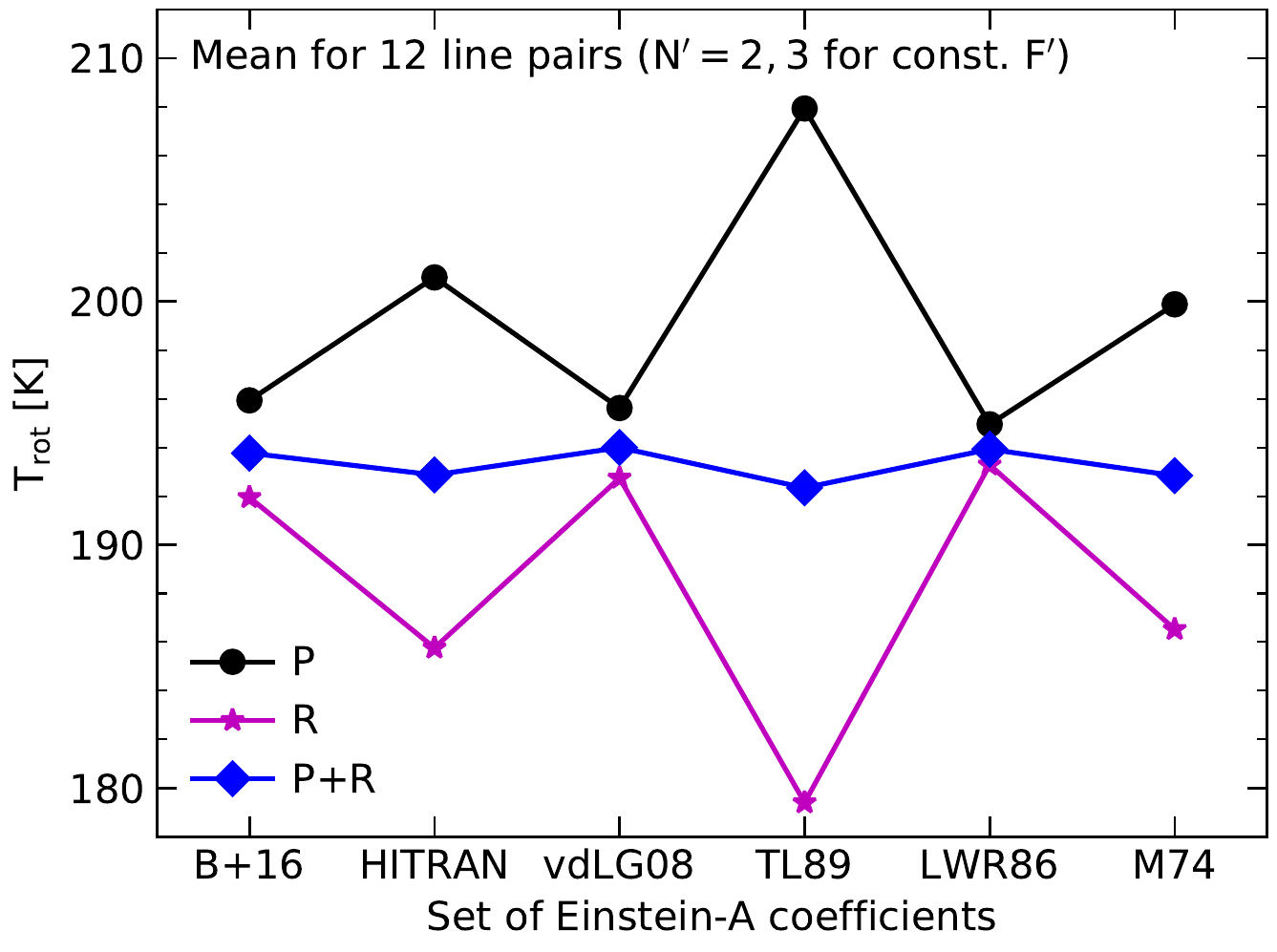}
\caption{Rotational temperature $T_\mathrm{rot}$ in kelvins for six sets of
  Einstein-A coefficients (see Fig.~\ref{fig:y_Ep}). The plotted mean
  temperatures were derived from 12 pairs of OH lines with fixed electronic
  substate $F^{\prime}$ originating from the second and third rotational upper
  level $N^{\prime}$. The branch was either P (circles) or R (stars). The
  diamonds show the resulting $T_\mathrm{rot}$ for a combination of the
  $y(E^{\prime})$ regression slopes for both branches (mean of inverse
  $T_\mathrm{rot}$).}
\label{fig:Trot_Aset}
\end{figure}

The differences in the dependence of the P- and R-branch-based populations on
$E^{\prime}$ for the investigated sets of Einstein-A coefficients imply
deviations in the related rotational temperatures
\begin{equation}\label{eq:Trot}
  T_\mathrm{rot} = - \frac{1}{k_\mathrm{B}
                  \frac{\mathrm{d}y}{\mathrm{d}E^{\prime}}} 
\end{equation}
\citep{mies74,noll18b}, where $k_\mathrm{B}$ is the Boltzmann constant and
${\mathrm{d}y}/{\mathrm{d}E^{\prime}}$ represents the slope of a regression
line in a $y(E^{\prime})$ plot like Fig.~\ref{fig:y_Ep} for the included
level populations. For a quantitative $T_\mathrm{rot}$ comparison, we
considered pairs of levels with a difference in $N^{\prime}$ of 1 where the
populations were derived from reliable lines (class 3) of the same OH band,
$F^{\prime}$, and branch. Only those pairs that are available for the P and R
branches were selected. This resulted in 35 pairs that are covered by all
sets of Einstein-A coefficients up to $N^{\prime} = 12$. The sample is
relatively small since R-branch lines are often blended. The highest number
of 12 pairs is found for the combination of the lowest $N^{\prime}$ of 2 and 3
($N^{\prime} = 1$ does not exist for the R branch). Mean results for these 12
pairs (which minimise the measurement uncertainties) are shown in
Fig.~\ref{fig:Trot_Aset}. The $T_\mathrm{rot}$ differences based on higher
$N^{\prime}$ agree qualitatively.

For the P branch, Fig.~\ref{fig:Trot_Aset} reveals a wide range of mean
temperatures between 195\,K for LWR86 and 208\,K for TL89, i.e. the selection
of the set of Einstein-A coefficients strongly affects the derivation of
absolute $T_\mathrm{rot}$. The situation is better if the extremely high value
for TL89 is neglected. In this case, the maximum difference (now limited by
the HITRAN-related result) is only 6\,K instead of 13\,K. Moreover, the
$T_\mathrm{rot}$ for the two most recent sets \mbox{B+16} and vdLG08 agree well
with the minimum related to LWR86. The temperature differences are consistent
with those derived by \citet{liu15} for low-$N^{\prime}$
P$_1$-branch lines of the OH bands \mbox{(3-0)}, \mbox{(5-1)}, \mbox{(6-2)},
\mbox{(8-3)}, and \mbox{(9-4)} based on observations with a Czerny--Turner
spectrometer at Xinglong in China. The differences between the highest and
lowest $T_\mathrm{rot}$ related to TL89 and LWR86, respectively (\mbox{B+16}
was not published yet), were between 9\,K for \mbox{OH(3-0)} and 17\,K for
\mbox{OH(8-3)} with the same mean of 13\,K. The trend of decreasing
$T_\mathrm{rot}$ differences for OH bands with longer central wavelengths can
also be observed in our data. Taking the differences between HITRAN and
\mbox{B+16} as an example, we find between 2\,K for \mbox{OH(3-0)}P$_1$ and
9\,K for \mbox{OH(6-1)}P$_2$ for the 12 selected line combinations. In this
context, the result of \citet{hart19b} for the P$_1$ branch of \mbox{OH(4-2)}
is interesting. Based on data from an astronomical spectrograph at Apache
Point in the USA, he found a maximum difference of 3\,K for the same five sets
investigated by \citet{liu15}. If the minimum related to LWR86 is excluded,
the variation is only about 1\,K with the lowest $T_\mathrm{rot}$ related to
TL89.

Figure~\ref{fig:Trot_Aset} also shows $T_\mathrm{rot}$ based on R-branch
lines, which were not used in the discussed studies. The set-dependent results
are remarkable since they mirror those for the P-branch lines. Now,
$T_\mathrm{rot}$ ranges from 179\,K for TL89 to 193\,K for LWR86, i.e. the
maximum difference of 14\,K is very similar to the result for the P branch but
the sign is reversed. Moreover, all $T_\mathrm{rot}$ related to the R branch
are lower than those related to the P branch. Hence, the $T_\mathrm{rot}$
difference between P and R branch is between 2\,K for LWR86 and 29\,K for
TL89. For individual double pairs of lines, R-branch-related $T_\mathrm{rot}$
can also be higher than those for the P branch, i.e. LWR86 might not show the
smallest differences. However, the large discrepancies for TL89 are obvious in
any case.

As the P- and R-branch $T_\mathrm{rot}$ show an oppositional behaviour, we
averaged the slopes ${\mathrm{d}y}/{\mathrm{d}E^{\prime}}$ for both branches to
derive more robust temperatures. As demonstrated by Fig.~\ref{fig:Trot_Aset},
this was achieved. The mean value for all sets is 193.3\,K with a standard
deviation of only 1.0\,K. The latter represents less than 20\% of the
variation for the individual branches. Consequently, the combination of P-
and R-branch data can significantly reduce the impact of the choice of the
Einstein-A coefficients on the quality of the resulting $T_\mathrm{rot}$.
However, in practice, this will be difficult to apply due to the difficulties
in measuring R-branch lines at moderate spectral resolution. Hence, it is more
promising to improve the Einstein-A coefficients by a better handling of the
vibration--rotation interaction \citep{pendleton02}, which appears to be the
main reason for the set-dependent $T_\mathrm{rot}$ discrepancies. Data as
plotted in Fig.~\ref{fig:Trot_Aset} can provide important constraints for this
purpose. 

\begin{figure*}[t]
\includegraphics[width=17cm]{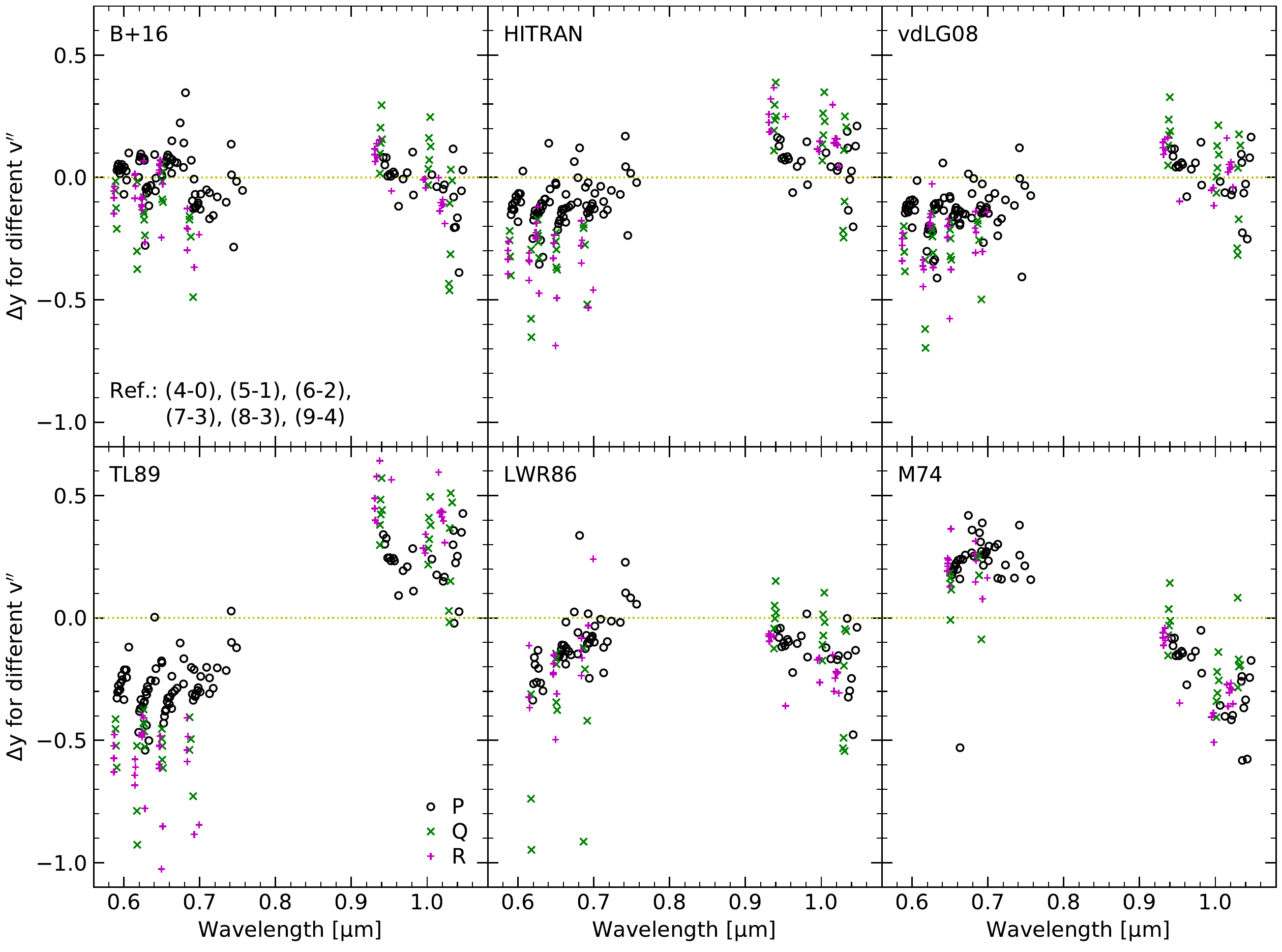}
\caption{Difference in logarithmic OH level populations $\Delta y$ by the
  change of the lower vibrational level $v^{\prime\prime}$ of the measured
  transitions as a function of the mean wavelength of the $\Lambda$ doublet in
  micrometres for six sets of Einstein-A coefficients. For the reference bands
  \mbox{OH(4-0)}, \mbox{OH(5-1)}, \mbox{OH(6-2)}, \mbox{OH(7-3)},
  \mbox{OH(8-3)}, and \mbox{OH(9-4)}, the resulting $\Delta y$ are zero and
  are therefore not shown. \mbox{OH(3-0)} is also neglected since it is the
  only band with $v^{\prime} = 3$. Finally, additional bands at short
  wavelengths are missing in the case of the limited sets of
  \citet{langhoff86} and \citet{mies74}. For more details, see legend and
  Fig.~\ref{fig:y_Ep}.}
\label{fig:dy_lam_dvpp}
\end{figure*}

Another population-independent evaluation of Einstein-A coefficients is
possible for transitions with the same upper and lower levels except for a
different $v^{\prime\prime}$. For the comparison of the related $y$, it was
necessary to define a reference OH band for each $v^{\prime}$ between 4 and 9,
where we have line measurements for two or more bands. We preferentially
selected bands with good quality data in the middle of the covered wavelength
range: \mbox{OH(4-0)}, \mbox{OH(5-1)}, \mbox{OH(6-2)}, \mbox{OH(7-3)},
\mbox{OH(8-3)}, and \mbox{OH(9-4)}. The resulting $\Delta y$ are plotted in
Fig.~\ref{fig:dy_lam_dvpp} as a function of line wavelength for the six sets
of Einstein-A coefficients. In the case of \mbox{B+16}, population ratios for
182 pairs of reliable $\Lambda$ doublets (class 3) are shown. For LWR86 and
M74, this number is only 136 due to the limitations in $N^{\prime}$ and
$\Delta v$ (Sect.~\ref{sec:fullpop}). The plots indicate a complex behaviour
where the $\Delta y$ depend on band, branch, and $N^{\prime}$ in a different way
for each set of Einstein-A coefficients. The data points for the lowest
$N^{\prime}$, which tend to cluster for each band, show a clear trend with
wavelength (or $\Delta v$) for all sets except \mbox{B+16}. The data for
HITRAN, vdLG08, and TL89 indicate an increase of $\Delta y$ with wavelength,
whereas the M74 data show a decrease. For LWR86, $\Delta y$ is mainly
negative, i.e. the reference bands in the middle of the wavelength range with
$\Delta y = 0$ (not plotted) indicate the highest relative populations.

The overall performance of each set can be evaluated by measuring the mean
absolute $\Delta y$ for line pairs where Einstein-A coefficients are
available in all sets. The corresponding results for 127 line pairs fulfilling
$\Delta v \le 5$ and $N^{\prime} \le 12$ are provided in
Table~\ref{tab:compAset}. The highest and hence worst
$\langle |\Delta y| \rangle$ were found for M74 (0.41) and TL89 (0.38). Lower
but still unsatisfactory values of around 0.18 were obtained for HITRAN,
vdLG08, and LWR86. \mbox{B+16} clearly shows the best performance with a value
of 0.11. Table~\ref{tab:compAset} also contains $\langle |\Delta y| \rangle$
depending on branch. The best results are obtained for the P branch for all
sets except M74, which is unsatisfactory for all branches. 

The large $\Delta y$ and their trend with wavelength for M74 and TL89 shown in
Fig.~\ref{fig:dy_lam_dvpp} were already found by \citet{cosby07}. Also using
UVES data, they compared the populations derived from the P$_1$(1) line of the
accessible OH bands with $v^{\prime}$ of 6, 8 and 9. Including the transition
probabilities of M74, LWR86, TL89, and \citet{goldman98}, their analysis
favoured the latter, i.e. the main input source for HITRAN. \citet{cosby07}
explained the bad performance of the TL89 coefficients by the erroneous
intensity calibration of data used for the applied empirical dipole moment
function (DMF), which is the basis for the calculation of the transition
probabilities. Population comparisons for OH lines from near-infrared bands
with low $\Delta v$ mostly not covered by UVES were performed by
\citet{oliva13} based on observations between 0.95 and 2.4\,$\mu$m with the
high-resolution echelle spectrograph GIANO at the Telescopio Nazionale Galileo
at the La Palma Observatory in Spain. The results show clear discrepancies
between populations derived from lines of bands with $\Delta v = 2$, 3 and 4
for the Einstein-A coefficients from \citet{vanderloo07}. Interestingly, the
corresponding trend with wavelength displayed in Fig.~\ref{fig:dy_lam_dvpp}
seems to be reversed for bands at longer wavelengths. In general, it can be
expected that the accuracy of Einstein-A coefficients for bands with high
$v^{\prime}$ in the optical tends to be worse than in the case of bands with
low $v^{\prime}$ in the near-infrared. Theoretical ab initio DMF calculations
as used by \citet{mies74} and \citet{vanderloo07,vanderloo08} are more
uncertain for internuclear distances between the O and H atom that are far
from the equilibrium. Moreover, the input data for empirical DMFs
\citep{turnbull88,turnbull89,nelson90}, theoretically extended empirical DMFs
\citep{goldman98}, and modified ab initio DMFs \citep{langhoff86,brooke16}
were mainly restricted to low $v$ or low $\Delta v$. 

\begin{figure*}[t]
\includegraphics[width=17cm]{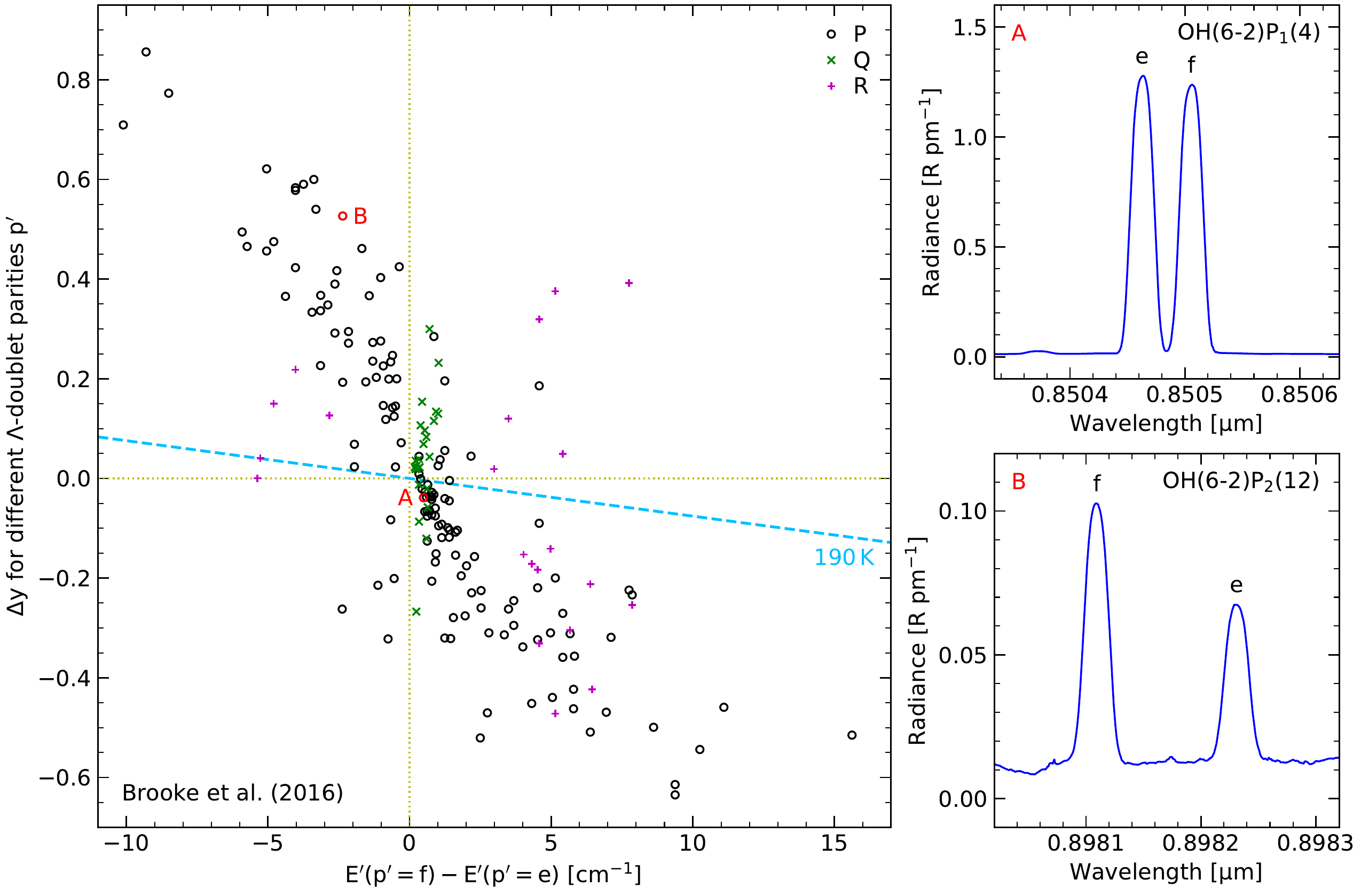}
\caption{Difference in logarithmic OH level populations $\Delta y$ of the
  separately measured components of reliable $\Lambda$ doublets indicated by
  the upper level parity f and e as a function of the corresponding difference
  in the upper level energy $E^{\prime}$. The results for different branches
  (P, Q, and R) are marked by different symbols and colours (see legend).
  Examples for small and large population discrepancies are marked by the
  letters A and B. On the right-hand side, the corresponding $\Lambda$-doublet
  spectra are plotted with indicated line identifications. The dashed line in
  the main plot indicates the effect of a thermal population with a
  temperature of 190\,K on $\Delta y$.}
\label{fig:dy_dEp_lambda}
\end{figure*}

As discussed in Sect.~\ref{sec:lineint}, about half of the measured $\Lambda$
doublets are resolved due to the high spectral resolving power of UVES. This
allowed us to systematically study deviations between the Einstein-A
coefficients of the e and f components. The older sets of transition
probabilities (M74, LWR86, and TL89) do not provide information on the
individual components. The HITRAN database \citep{gordon17} contains these
components but the Einstein-A coefficients were just set to the value of the
corresponding doublet. Finally, vdLG08 and \mbox{B+16} consider $\Lambda$
doubling but the differences between the coefficients are very small. For
\mbox{B+16}, the mean relative difference for our sample of 723 doublets is
only 0.04\%. The largest deviations are related to P- and Q-branch lines with
high $N^{\prime}$. The maximum in our sample of 0.25\% is linked to
\mbox{OH(5-1)Q$_2$(6)}. As the corresponding values for vdLG08 are almost
identical, it is sufficient to use only \mbox{B+16} coefficients for the
comparison of the $\Lambda$-doublet components.

Figure~\ref{fig:dy_dEp_lambda} shows the results for 185 reliable (class 3)
doublets with resolved components. The logarithmic population ratio $\Delta y$
for f minus e for the upper state parity $p^{\prime}$ is plotted as a function
of the corresponding difference in $E^{\prime}$. The latter is negative for
$F^{\prime} = 2$ lines of the P and R branch according to the parity
definition used by \citet{brooke16}. If the theoretically predicted equality
of the transition probabilities was true, $\Delta y$ should be close to 0.
Small deviations in the populations are possible due to the small differences
in $E^{\prime}$. Assuming a Boltzmann-like distribution for a typical kinetic
temperature of 190\,K at altitudes of the OH emission layer at Cerro Paranal
\citep{noll16}, there would be $\Delta y = +0.08$ for
$\Delta E^{\prime} = -10$\,cm$^{-1}$ and the same amount with negative sign
for the corresponding positive energy difference. However, the true $\Delta y$
are about 1 order of magnitude larger. The average absolute discrepancy in
$\Delta y$ is 0.22. In the case of large energy differences of at least
5\,cm$^{-1}$, it would even be 0.37, which corresponds to a ratio of 1.4. The
clear differences in the strengths of the $\Lambda$-doublet components are
also illustrated by two example spectra. The weakly separated
\mbox{OH(6-2)P$_1$(4)} doublet already shows slight differences in the
intensity ($\Delta y = -0.04$). For the widely separated
\mbox{OH(6-2)P$_2$(12)} pair, the f component is about 1.7 times brighter than
the e component ($\Delta y = 0.53$). It is astonishing that it appears that
these large effects have not been recognised, so far. They can only be
explained by inadequate Einstein-A coefficients since the P, Q, and R branches
behave differently. The $\Delta y$ related to P-branch lines could be fitted
by a non-LTE Boltzmann-like distribution with about 20\,K. However, the
$\Delta y$ distributions for the Q and R branches are less clear. A convincing
regression line cannot be drawn, and even if the fitting was performed, the
slopes would be very different. This rules out a significant impact of
possible non-LTE-inducing propensity differences for the population of the e
and f states, even if the parity definitions are changed or more complex
dependencies involving several level parameters are considered.

\subsection{Correction of Brooke et al. coefficients}\label{sec:corrcoeff}

The discussion in the previous section has shown that the currently available
sets of Einstein-A coefficients are not satisfactory, especially with respect
to Q-branch lines and $\Lambda$-doublet components. Overall, the \mbox{B+16}
set is the most promising since it is the most complete in terms of the
included lines, shows the smallest band-to-band variations for constant
$v^{\prime}$, and is only slightly worse than LWR86 with respect to the
population deviations between different branches. Hence, we focus on the
\mbox{B+16} set for the rest of this paper. However, as the remaining issues
can still negatively affect the evaluation of OH level population
distributions as shown in Fig.~\ref{fig:y_Ep}, we tried to improve the
coefficients to result in more consistent population ratios in the diagnostic
plots discussed in Sect.~\ref{sec:comparisons}. Our approach is fully
empirical, i.e. it is based on regression lines and correction factors, and
consists of three steps related to the correction of the discrepancies
revealed by Figs.~\ref{fig:dy_dEp_branch}, \ref{fig:dy_lam_dvpp}, and
\ref{fig:dy_dEp_lambda}. Complex fitting approaches involving theoretical
DMF-based calculations of the Einstein-A coefficients are out of the scope of
this study. 

\begin{figure}[t]
\includegraphics[width=8.3cm]{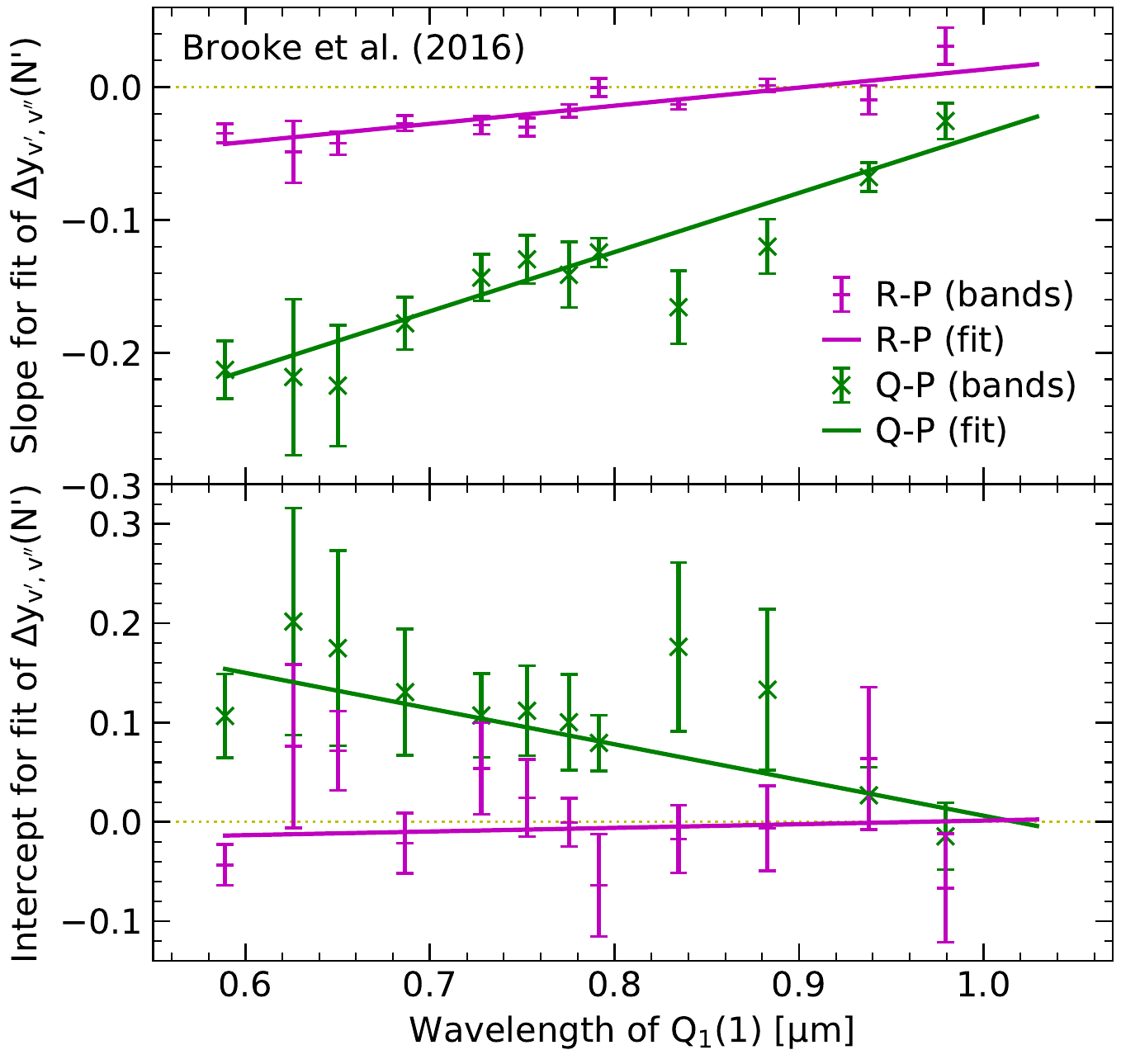}
\caption{Fit of the differences in logarithmic OH level populations for Q
  (crosses) and R branches (horizontal bars) relative to the P branch as a
  function of the upper rotational level $N^{\prime}$ for each band where a
  reliable linear regression could be performed. The symbols with error bars
  show the resulting slope (upper panel) and intercept (lower panel) and their
  respective uncertainties. These values based on Einstein-A coefficients of
  \citet{brooke16} were also fitted by performing error-weighted linear fits
  depending on the wavelength of Q$_1$(1) for each band. The fit lines are
  displayed.}
\label{fig:mc_lam_branch}
\end{figure}

We started with a correction of the discrepancies between the populations
derived from different branches as shown in Fig.~\ref{fig:dy_dEp_branch}. For
this purpose, we used 184 suitable, highly reliable $\Lambda$ doublets
classified as 33, i.e. with a primary and secondary class of 3
(Sect.~\ref{sec:lineint}). It turned out that a linear fit of $\Delta y$ for
Q versus P and R versus P shows the best performance for $N^{\prime}$ as the
independent variable (instead of the plotted $\Delta E^{\prime}$) and a
separate fit for each OH band. The resulting slopes and intercepts and their
uncertainties are provided in Fig.~\ref{fig:mc_lam_branch}. There are no data
points for \mbox{OH(5-0)}, \mbox{OH(9-5)}, and \mbox{OH(4-1)} due to
relatively high uncertainties caused by an insufficient number of reliable
line measurements. For the remaining 12 bands, 5 to 16 (3 to 8) $\Lambda$
doublets could be used for the fit of the difference between R and P branch
(Q and P branch). The slopes for the Q branch are clearly more negative than
those for the R branch. The mean values are $-0.15$ and $-0.02$, i.e. the
apparent populations decrease by 14 and 2\% compared to the P branch for an
increase of $N^{\prime}$ by 1. The corresponding mean intercepts are $+0.11$
and $+0.01$, i.e. nearly zero for R versus P. The data points are plotted as a
function of the central wavelength of the Q$_1$(1) doublet of each band as
especially the slope indicates a significant trend with wavelength, which
suggests that the branch-related errors in the \mbox{B+16} Einstein-A
coefficients depend on the energy difference between the upper and lower
state. The discrepancies between the branches appear to be smaller for longer
wavelengths and might even vanish around 1\,$\mu$m. We used these correlations
for a more robust correction approach involving error-weighted linear fits of
slope and intercept as a function of Q$_1$(1) wavelength. The error weighting
allowed us to consider the strong dependence of the uncertainties on the band.
The resulting fits are also shown in Fig.~\ref{fig:mc_lam_branch}. The slopes
change with $+0.45 \pm 0.05$ (Q) and $+0.14 \pm 0.03$ (R) per micrometre. For
the intercepts, we obtain changes of $-0.36 \pm 0.08$ (Q) and
$+0.04 \pm 0.11$ (R) per micrometre.

\begin{figure*}[t]
\includegraphics[width=17cm]{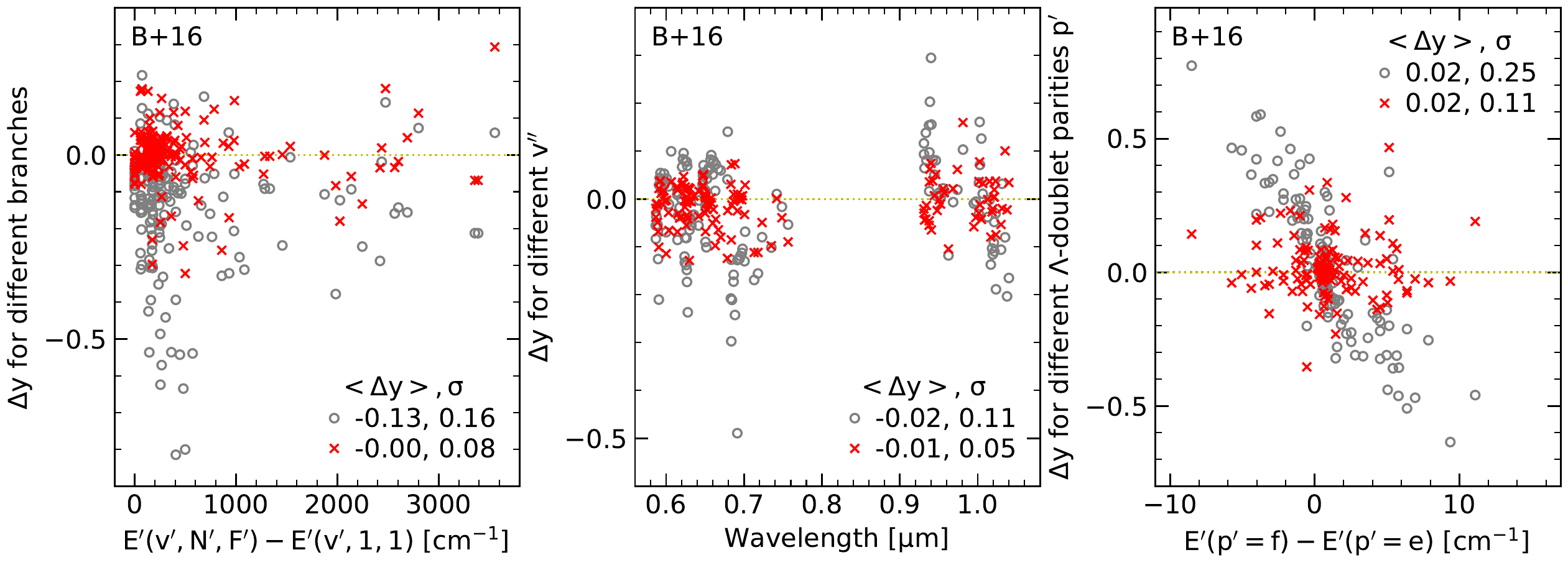}
\caption{Correction of \citet{brooke16} Einstein-A coefficients. The measured
  (circles) and corrected (crosses) OH level population ratios $\Delta y$ are
  given for changes in the branch (left; c.f. Fig.~\ref{fig:dy_dEp_branch}),
  lower vibrational level $v^{\prime\prime}$ (middle; c.f.
  Fig.~\ref{fig:dy_lam_dvpp}), and the $\Lambda$-doublet component (right;
  c.f. Fig.~\ref{fig:dy_dEp_lambda}). Each plot lists the mean $\Delta y$ and
  standard deviation $\sigma$ for the measured and corrected data.}
\label{fig:dycorr}
\end{figure*}

According to Eq.~(\ref{eq:ydef}), the reciprocals of the population ratios
$\Delta y$ from our two-step fitting procedure indicate the systematic
deviations of the Einstein-A coefficients. Thanks to the derived regression
lines, they can also be predicted for missing lines and bands, at least in the
covered wavelength range. For the correction, we need to define a reference.
While the Q-branch coefficients do not appear to be reliable in general, it
would be arbitrary to choose the P or the R branch. However,
Fig.~\ref{fig:Trot_Aset} suggests that the errors in the Einstein-A
coefficients are minimised if both branches are combined with the same weight.
Hence, we corrected the transition probabilities for the three branches by
using half the fitted deviation between P and R branch as the reference.
Figure~\ref{fig:dycorr} (left) shows $\Delta y$ for the 184 considered pairs
of $\Lambda$ doublets before and after the correction. While in the former
case the mean value and standard deviation are $-0.13$ and $0.16$, the
corrected data reveal $0.00$ and $0.08$. Thus, the offsets vanished completely
on average and the scatter was reduced by a factor of 2. 

In the next step, we corrected $\Delta y$ offsets between lines differing only
in $v^{\prime\prime}$ as shown in Fig.~\ref{fig:dy_lam_dvpp}. For this purpose,
we focused on relatively bright lines with $N^{\prime} \le 4$ for the P and R
branch and $N^{\prime} = 1$ for the Q branch. As these $\Lambda$ doublets
indicate similar $\Delta y$ (scatter of 0.03), differences in the selected
line subsets do not critically affect the mean values. For each considered OH
band, 8 to 14 reliable pairs of $\Lambda$ doublets of class 33 were available
(89 in total). For the correction of $\Delta y$ by changing the Einstein-A
coefficients, we used the same reference bands for each $v^{\prime}$ as
discussed in Sect.~\ref{sec:comparisons}. The choice is motivated by the
accessibility and quality of the line measurements with UVES. It does not
necessarily include the bands with the most realistic transition
probabilities. As discussed in Sect.~\ref{sec:comparisons}, coefficients of
bands with low $v^{\prime}$ and $\Delta v$ tend to be more reliable as the DMF
calculations are less challenging and the experimental data are more abundant.
Our reference bands have $v^{\prime}$ between 3 and 9 and $\Delta v$ between 3
and 5. Hence, the most promising bands are beyond the UVES wavelength range.
Nevertheless, there does not appear to be a strong quality gradient with
$\Delta v$ or wavelength ($v^{\prime}$ cannot be tested) since the \mbox{B+16}
coefficients do not show such a dependence of $\Delta y$ for the covered bands
in Fig.~\ref{fig:dy_lam_dvpp} (in contrast to the other investigated sets).
Note that this is different from the situation for the branches illustrated in
Fig.~\ref{fig:mc_lam_branch}. In the end, we shifted the mean $\Delta y$ for
eight OH bands to zero by multiplying the Einstein-A coefficients by factors
between 0.85 for \mbox{OH(7-2)} and 1.06 for \mbox{OH(8-4)}. This reduces the
scatter in the measured populations for fixed $v^{\prime}$ in any case, even if
the choice of the reference bands might not be optimal.
Figure~\ref{fig:dycorr} (middle) shows the corresponding results for 143
$\Lambda$ doublets. The discussed corrections (also including the
branch-related modifications) change the mean $\Delta y$ and standard
deviation from $-0.02$ and $0.11$ to $-0.01$ and $0.05$.

Finally, we corrected the $\Delta y$ between the $\Lambda$-doublet components
as shown in Fig.~\ref{fig:dy_dEp_lambda}. This is necessary in order to also
use doublets with only one reliable component for the study of the OH level
populations discussed in Sect.~\ref{sec:ohpop}. For the change of the
Einstein-A coefficients, we assumed a natural population discrepancy between
the two components consistent with a temperature of 190\,K as illustrated in
Fig.~\ref{fig:dy_dEp_lambda}. We fitted the remaining $\Delta y$ for each
branch using $N^{\prime}$ as the independent variable due to a better
performance with respect to linear regressions compared to $\Delta E^{\prime}$.
This approach requires us to flip the sign for the data points with negative
$\Delta E^{\prime}$. This is reasonable since the amount of the deviations is
very similar for $\Lambda$ doublets with $F^{\prime}$ of 1 and 2. For the fits
related to P, Q, and R, we considered 99, 18, and 11 resolved doublets of
class 33. The resulting slopes (and intercepts) are $-0.035 \pm 0.003$
($+0.13 \pm 0.03$), $+0.065 \pm 0.019$ ($-0.19 \pm 0.08$), and
$-0.035 \pm 0.035$ ($+0.43 \pm 0.48$), respectively. Hence, the effect for the
Q branch seems to be twice as large as for the P branch and to also have a
different sign. The R branch might behave similarly to the P branch but the
uncertainties are high. Assuming that the e and f components equally
contribute to the fitted differences, we corrected the Einstein-A
coefficients for the reliable $N^{\prime}$ range of the fits, i.e. we neglected
doublets with low $N^{\prime}$ where both components are not sufficiently
separated or the fit crossed $\Delta y = 0$. Figure~\ref{fig:dycorr} (right)
shows the resulting change in the mean $\Delta y$ and scatter for the
investigated 128 doublets. While the small mean value of $+0.02$ did not
significantly change, the standard deviation was clearly reduced from $0.25$
to $0.11$.

\section{OH level populations}\label{sec:ohpop}

\subsection{Mean populations and rotational temperatures}\label{sec:poptrot}

\begin{figure*}[t]
\includegraphics[width=17cm]{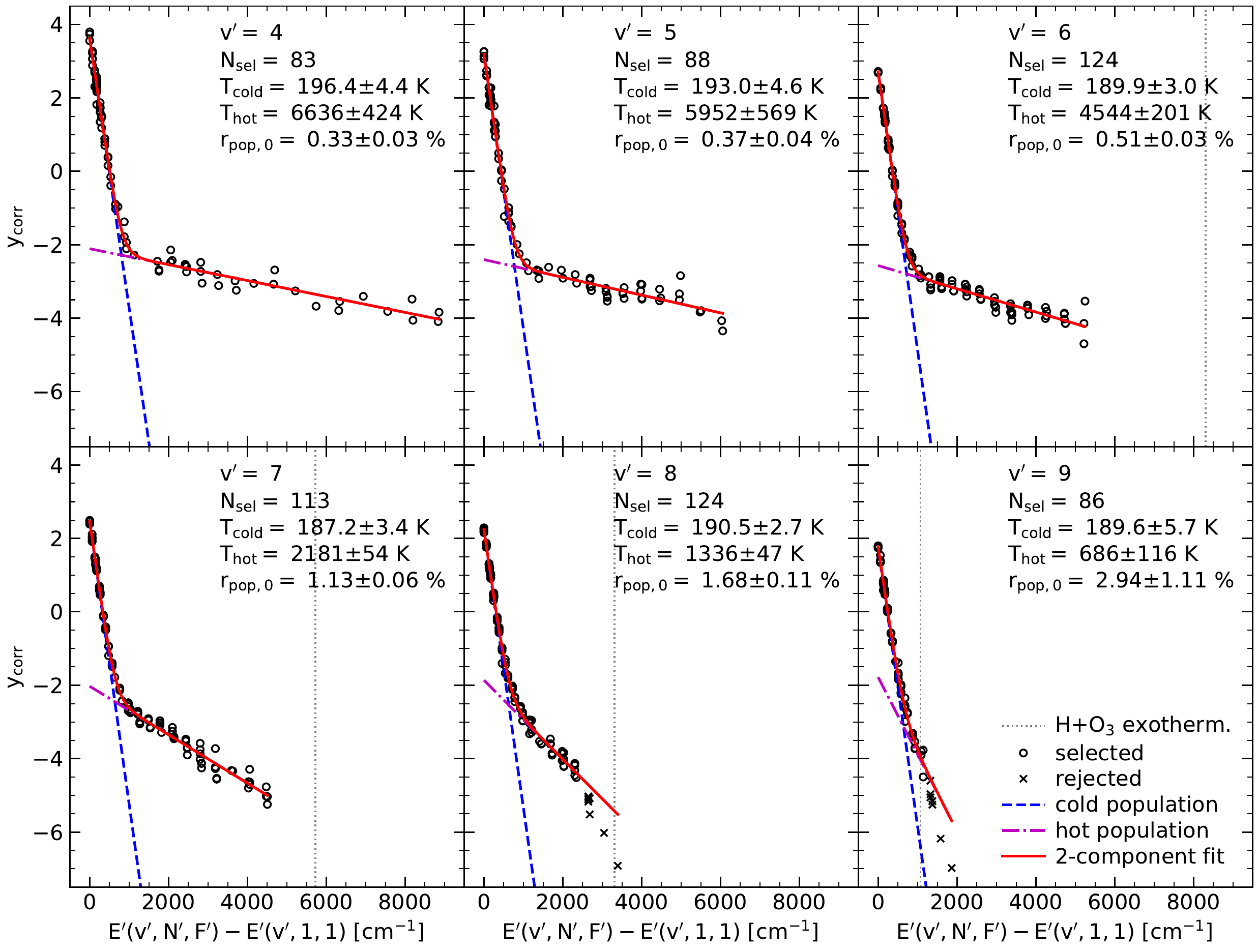}
\caption{Distribution of logarithmic OH level populations for the corrected
  \citet{brooke16} Einstein-A coefficients. Except for \mbox{OH(7-1)}, all
  $\Lambda$ doublets with at least one reliable component are considered. The
  populations are plotted in separate panels for each upper vibrational level
  $v^{\prime}$ between 4 and 9. The given level energies in inverse centimetres
  are relative to the lowest $v^{\prime}$-related energy. The populations can
  be fitted by means of a two-component rotational temperature fit. All data
  points involved are marked by circles. The few exceptions (indicated by
  crosses) are close to or above the exothermicity limit of the
  hydrogen--ozone reaction, which is marked by vertical dotted lines. For
  these data, a more complex fit would be necessary. The final fit curves are
  displayed by solid lines. The underlying cold and hot components are shown
  by dashed and dot-dashed lines, respectively. Each panel lists the
  $v^{\prime}$-specific number of selected data points $N_\mathrm{sel}$, the
  rotational temperatures of the cold and hot components $T_\mathrm{cold}$ and
  $T_\mathrm{hot}$ in kelvins, and the ratio $r_\mathrm{pop,0}$ of the
  populations of both linear fit components for the lowest $v^{\prime}$-related
  energy in per cent.}
\label{fig:ycorr_dEp}
\end{figure*}

With the correction of the \mbox{B+16} Einstein-A coefficients in
Sect.~\ref{sec:corrcoeff}, we minimise the scatter in the OH level
populations for upper states with different measured lines. Moreover, the
change of the populations with $N^{\prime}$ appears to be more reliable due to
the promising combination of P- and R-branch data for $T_\mathrm{rot}$
estimates (Fig.~\ref{fig:Trot_Aset}). The resulting population distributions
for $v^{\prime}$ between 4 and 9 are shown in Fig.~\ref{fig:ycorr_dEp}. We
neglect $v^{\prime} = 3$ due to the lack of high-$N^{\prime}$ states in the UVES
data (Fig.~\ref{fig:Ep_vp}), which does not allow us to describe this
population distribution in detail. As already briefly discussed in
Sects.~\ref{sec:intro} and \ref{sec:fullpop}, there is a characteristic
pattern for increasing level energy with a steep population decrease for low
$N^{\prime}$ and a rather slow decrease for high $N^{\prime}$. The difference
between high and low $N^{\prime}$ tends to increase with decreasing $v^{\prime}$.
Moreover, the change between the two extremes does not appear to happen
continuously with increasing $N^{\prime}$. Instead, the transition is mostly
localised in a narrow $\Delta E^{\prime}$ interval of a few hundred inverse
centimetres. This known pattern \citep{cosby07,oliva15} suggests the
definition of a cold and a hot population for each $v^{\prime}$, which can be
described by corresponding $T_\mathrm{rot}$ and population ratios for fixed
level energies \citep{oliva15,kalogerakis18,kalogerakis19}.

We applied this concept by fitting the natural logarithm of the sum of two
exponential Boltzmann terms as a function of $\Delta E^{\prime}$ ($E^{\prime}$
relative to the energy for $N^{\prime} = 1$ and $F^{\prime} = 1$) to the
corrected $y$ for each $v^{\prime}$. We considered the populations from all
$\Lambda$ doublets with quality classes above 0. For classes 1 and 2, we
derived the doublet-related populations from the reliable components. An
inspection of the change of the populations with increasing $\Delta E^{\prime}$
resulted in the rejection of the highest $N^{\prime}$ levels for $v^{\prime} = 8$
and 9 as the related populations cannot be reproduced satisfactorily by a
two-component fit. In the case of $v^{\prime} = 9$, the seven measurements with
$N^{\prime} \ge 10$ were neglected. All corresponding $E^{\prime}$ are between
250 and 790\,cm$^{-1}$ above the exothermicity limit of the hydrogen--ozone
reaction (Sect.~\ref{sec:lineint}), which can explain the rapid population
decrease in this energy range, which could not clearly be constrained before
due to a lack of data \citep{cosby07,noll18b}. In the case of $v^{\prime} = 8$,
a strong decrease of the populations is found for eight measurements related
to $N^{\prime} \ge 14$ with $E^{\prime}$ between 660\,cm$^{-1}$ below and
90\,cm$^{-1}$ above the exothermicity limit. This is an interesting result as
it provides valuable constraints on the nascent populations and the relaxation
process from $v^{\prime} = 9$ to 8. The drop of the populations below the
exothermicity limit also seems to be present in the population distribution
of \citet{cosby07}, also based on UVES spectra \citep{hanuschik03,cosby06}.
However, the authors do not discuss this phenomenon. Our $v^{\prime} \le 7$
data, which are related to energies of more than 1,200\,cm$^{-1}$ below the
limit, do not show a cut in the populations. The data of \citet{cosby07} are
not conclusive here either.

The remaining population measurements for each $v^{\prime}$, which varied
between 83 for $v^{\prime} = 4$ and 124 for $v^{\prime} = 6$ and 8, were fitted
with our two-component model by means of robust least-squares minimisation,
which resulted in the same best fits for a wide range of start values.
Figure~\ref{fig:ycorr_dEp} shows the final best fits and also indicates the
corresponding temperatures $T_\mathrm{cold}$ and $T_\mathrm{hot}$ as well as the
ratio of the hot and cold populations for $\Delta E^{\prime} = 0$,
$r_\mathrm{pop,0}$. Under consideration of the fit uncertainties, the best-fit
$T_\mathrm{cold}$ are very similar and consistent with a temperature of 190\,K,
i.e. the typical ambient temperature at OH emission altitudes \citep{noll16}.
Only $v^{\prime} = 4$ with $196 \pm 4$\,K might slightly be higher. This could
point to the weak influence of an intermediate population for low $v^{\prime}$.
The second highest $T_\mathrm{cold}$ value of $193 \pm 5$\,K for
$v^{\prime} = 5$ would be in agreement with this interpretation. Note that
fixing the fit to a $T_\mathrm{cold}$ of 190\,K did not significantly change
the other parameters. The differences were much smaller than the
uncertainties. In contrast to $T_\mathrm{cold}$, $T_\mathrm{hot}$ shows a
strong trend with $v^{\prime}$. The temperatures increase from about 700\,K
for $v^{\prime} = 9$ to about 7,000\,K for $v^{\prime} = 4$. In parallel,
$r_\mathrm{pop,0}$ decreases from about 3\% for $v^{\prime} = 9$ to about
0.3\% for $v^{\prime} = 4$, i.e. hot populations with higher $T_\mathrm{hot}$
show lower contributions to the total population at low $N^{\prime}$. The
strong change in $r_\mathrm{pop,0}$ appears to be mainly caused by the
decrease of the cold population with increasing $v^{\prime}$ since the
$\Delta E^{\prime} = 0$ intercepts of the lines describing the hot populations
are located at similar $y$ values of around $-2$ in Fig.~\ref{fig:ycorr_dEp}.
The fits for $v^{\prime} \le 7$ (no rejection of states) are convincing with
respect to the assumption of a homogeneous hot population, which can be
described by a single temperature. Nevertheless, some fine structure might
exist as the comparison of the individual measurements and the fit lines
suggest, although the possible population deviations appear to be not larger
than 30\%, which is small compared to population changes of the order of a
magnitude in the $v^{\prime}$-dependent energy ranges most contributing to
$T_\mathrm{hot}$. Hence, the two-component fits are quite robust as the listed
errors show. The highest uncertainties are related to $v^{\prime} = 9$ since
the hot population is essentially constrained in an energy range of less than
300\,cm$^{-1}$, which includes 12 measurements with $N^{\prime}$ of 8 and 9.

Two-component fits were previously performed by \citet{oliva15} based on a
near-infrared GIANO spectrum with a resolving power of 32,000 taken during
2\,hours with the spectrograph directly pointing to the night sky at the La
Palma Observatory. The investigated lines belong to OH bands with low
$\Delta v$ and are complementary to those covered by our study. For the
calculation of the populations, \citet{oliva15} used the Einstein-A
coefficients from \citet{vanderloo07}. For the fits, $T_\mathrm{cold}$ was
fixed at 200\,K. The resulting $T_\mathrm{hot}$ and $r_\mathrm{pop,0}$ varied
from about 1,300\,K and 1.8\% for $v^{\prime} = 8$ to about 7,000\,K and 0.23\%
for $v^{\prime} = 4$. Although errors were not reported, these values are in
good agreement with our results provided in Fig.~\ref{fig:ycorr_dEp}. The fit
parameters for $v^{\prime} = 9$ are highly uncertain. However, \citet{oliva15}
succeeded in fitting the populations for $v^{\prime} = 2$ and 3, which show an
extension of the trend found for the higher $v^{\prime}$. For $v^{\prime} = 2$,
$T_\mathrm{hot}$ and $r_\mathrm{pop,0}$ resulted in 12,000\,K and 0.14\%,
respectively. The GIANO data were refitted by \citet{kalogerakis18} with
unconstrained $T_\mathrm{cold}$, which resulted in temperatures of about 190\,K
but with larger scatter than in our case. For $T_\mathrm{hot}$, the general
trend was the same but with a large step from 900\,K for $v^{\prime} = 8$ to
4,000\,K for $v^{\prime} = 7$, which disagrees with our findings. Population
ratios were not provided by \citet{kalogerakis18}. \citet{noll18b} already
published populations related to $v^{\prime} = 9$ and P-branch lines based on
the UVES data used in this study and \mbox{B+16} Einstein-A coefficients.
\citet{kalogerakis19} fitted these populations and found $T_\mathrm{cold}$ and
$T_\mathrm{hot}$ of about 180\,K and 500\,K, respectively. Both temperatures
are lower than our results, but less than 2 standard deviations. The fit of
\citet{kalogerakis19} based on fewer data points seems to be related to a
higher impact of the hot population at low $N^{\prime}$. Our results for
$T_\mathrm{hot}$ allow for an interesting comparison to the $T_\mathrm{rot}$ of
the nascent populations of $v^{\prime}$ between 7 and 9, which were derived by
\citet{llewellyn78} using laboratory data from \citet{charters71}. Our
best-fit $T_\mathrm{hot}$ of $690 \pm 120$, $1340 \pm 50$, and $2180 \pm 50$\,K
agree well with their $760 \pm 20$, $1230 \pm 30$, and $1940 \pm 200$\,K,
which implies that the OH relaxation processes do not appear to significantly
affect the hot populations of the highest $v^{\prime}$.

The previous discussion has shown that bimodality is a good concept for the
description of the population distributions for each $v^{\prime}$. Moreover,
the derived $T_\mathrm{cold}$ are close to the expected effective ambient
temperatures for the $v^{\prime}$-dependent OH emission layers \citep{noll16}.
The increasing trend of $T_\mathrm{rot}$ derived from the lines with the
lowest $N^{\prime}$ for increasing $v^{\prime}$ \citep{cosby07,noll15,noll17} is
not found in the best-fit $T_\mathrm{cold}$. Hence, our fits could be used to
estimate the non-LTE contributions to such $T_\mathrm{rot}$,
$\Delta T_\mathrm{NLTE}$, which are an issue for the use of $T_\mathrm{rot}$ as
indicators of the temperatures in the mesopause region. \citet{kalogerakis18}
and \citet{kalogerakis19} compared $T_\mathrm{cold}$ fits with $T_\mathrm{rot}$
from linear regressions for levels with $\Delta E^{\prime}$ lower than 500 and
250\,cm$^{-1}$, respectively. The results indicate higher $T_\mathrm{rot}$
than $T_\mathrm{cold}$ at least for the highest $v^{\prime}$ (order of 20\,K).
However, the uncertainties are large due to the strong impact of the line
selection \citep{noll15}, uncertainties in the line intensities (unclear for
the GIANO data), and the choice of the Einstein-A coefficients
(Fig.~\ref{fig:Trot_Aset}). Hence, we applied a different approach by
directly taking the two-component fit for the measurement of $T_\mathrm{rot}$.
For this purpose, we derived the populations related to the first three
$P_1$-branch lines, which are often taken for $T_\mathrm{rot}$ determinations
\citep[e.g.][]{schmidt13,noll16}, from the fit curve at the corresponding
$v^{\prime}$-dependent $\Delta E^{\prime}$. The related $T_\mathrm{rot}$ were then
calculated by a linear regression of the three $y$ for each $v^{\prime}$.
Finally, the resulting $\Delta T_\mathrm{NLTE}$ is just the difference between
$T_\mathrm{rot}$ and $T_\mathrm{cold}$. This method is very robust as it is fully
based on the two-component fit, which relies on a high number of population
measurements. Hence, uncertainties related to individual lines are negligible. 

\begin{figure}[t]
\includegraphics[width=8.3cm]{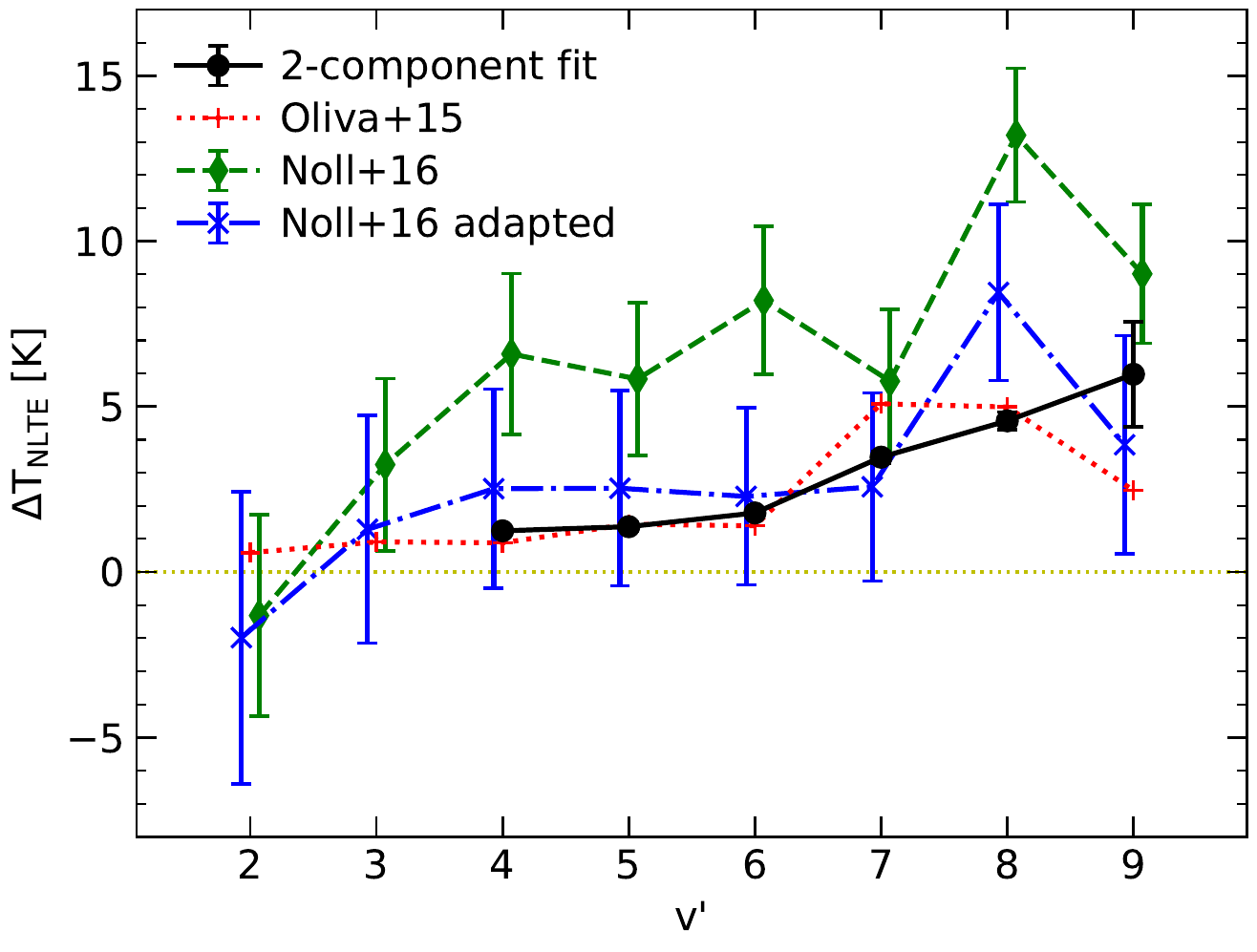}
\caption{Estimate of non-LTE contributions to rotational temperatures for the
  three lowest rotational lines with $F' = 1$ as a function of the upper
  vibrational level $v^{\prime}$. The differences of the $T_\mathrm{rot}$ for the
  two-component fits and the related $T_\mathrm{cold}$ shown in
  Fig.~\ref{fig:ycorr_dEp} are provided by circles and solid lines. Error
  bars (which are only larger than the symbols for $v^{\prime} = 9$) are also
  given. Two-component fits of OH level populations were also performed by
  \citet{oliva15} based on data from the near-infrared spectrograph GIANO at
  La Palma Observatory in Spain. We used their fit parameters (no errors) to
  calculate $\Delta T_\mathrm{NLTE}$ for $v^{\prime}$ from 2 to 9 (plus signs
  and dotted lines). For comparison, the figure also indicates the estimates
  by \citet{noll16} based on X-shooter spectra from Cerro Paranal and SABER
  data for a similar area (diamonds and dashes). As these
  $\Delta T_\mathrm{NLTE}$ were derived for Einstein-A coefficients from
  HITRAN, we recalculated the non-LTE effects using our corrected
  \citet{brooke16} coefficients (crosses and dot-dashed lines).}
\label{fig:delTnlte_vp}
\end{figure}

Our $\Delta T_\mathrm{NLTE}$ for $v^{\prime}$ between 4 and 9 are shown in
Fig.~\ref{fig:delTnlte_vp}. They increase relatively slowly between 4 and 6
from $1.2 \pm 0.1$\,K to $1.8 \pm 0.1$\,K and then faster to $6.0 \pm 1.6$\,K
for $v^{\prime} = 9$. As the latter value is relatively uncertain,
$4.6 \pm 0.3$\,K for $v^{\prime} = 8$ might also be the maximum deviation. The
errors were derived by displacing the hot component fit in both $y$ directions
according to the uncertainty in $r_\mathrm{pop,0}$ and refitting
$T_\mathrm{hot}$ as the only parameter to obtain modified
$\Delta T_\mathrm{NLTE}$. This approach considers that $r_\mathrm{pop,0}$ and
$T_\mathrm{hot}$ are anticorrelated and that the relative uncertainty in
$T_\mathrm{cold}$ is relatively small. Additional systematic uncertainties are
caused by assuming only two components. It is required that the fit line for
the hot component can be linearly extrapolated to $\Delta E^{\prime} = 0$. The
good quality of the fits in the transition region between the dominance of the
cold and hot components are promising. Nevertheless, contributions of
additional components of intermediate temperature cannot be excluded (as e.g.
for $v^{\prime} = 4$ due to a possibly elevated $T_\mathrm{cold}$). Fits with such
an additional component and a fixed $T_\mathrm{cold}$ of 190\,K, where the
best-fit parameters of the intermediate and hot populations are not well
constrained, showed possible $\Delta T_\mathrm{NLTE}$ increases by 10 to 30\%,
i.e. the significance of positive non-LTE effects for all $v^{\prime}$ would
remain high. It is hard to imagine situations where $\Delta T_\mathrm{NLTE}$
could significantly drop. A sharp cut of the hot population for low
$N^{\prime}$ would be inconsistent with a Boltzmann-like distribution as
expected for relaxation processes. Another source of possible systematic
errors are the Einstein-A coefficients, especially with respect to their
dependence on $N^{\prime}$. The latter was changed in
Sect.~\ref{sec:corrcoeff} for the \mbox{B+16} coefficients by the
branch-specific corrections. Hence, we tested what happens if we consider the
P- or R-branch data as the standards instead of a combination of both. These
modifications would change directly measured $T_\mathrm{rot}$ to values as
indicated in Fig.~\ref{fig:Trot_Aset}. However, the effect on the
two-component approach is much smaller. It is of the order of the already
small fit errors for all $v^{\prime}$. As expected, non-LTE contributions
related to R as the standard are lower than those for P. Furthermore, we
investigated the influence of the choice of the energy levels on
$\Delta T_\mathrm{NLTE}$ by also simulating line sets consisting of the first
two and first four P$_1$-branch lines. For $v^{\prime} = 8$ as an example,
these changes cause $\Delta T_\mathrm{NLTE}$ of 3.3 and 6.9\,K, which clearly
deviate from the plotted 4.6\,K. Hence, the non-LTE contributions are very
sensitive to the selected energy levels, which is consistent with the results
from \citet{noll15,noll18b}.

As shown in Fig.~\ref{fig:delTnlte_vp}, we also calculated
$\Delta T_\mathrm{NLTE}$ from the two-component fits of \citet{oliva15}.
Excluding their very uncertain fit parameters for $v^{\prime} = 9$, there is a
very good agreement with differences smaller than 0.4\,K. The only exception
is $v^{\prime} = 7$, where our non-LTE contributions are about 1.6\,K lower. As
the data basis and analysis were completely different (including different
Einstein-A coefficients), this convincing result demonstrates the robustness
of the approach. The GIANO-related data for $v^{\prime} = 2$ and 3 suggest that
$\Delta T_\mathrm{NLTE}$ decreases only very slowly with decreasing $v^{\prime}$.
The drop in $r_\mathrm{pop,0}$ seems to be nearly compensated by the increase
in $T_\mathrm{hot}$. 

Mean $\Delta T_\mathrm{NLTE}$ for $v^{\prime}$ from 2 to 9 at Cerro Paranal were
already derived by \citet{noll16} based on measurements of 25 OH bands and 2
O$_2$ bands (where non-LTE effects are less important) in optical and
near-infrared spectra from the echelle spectrograph X-shooter as well as from
OH emission and kinetic temperature profile measurements with SABER. The
$\Delta T_\mathrm{NLTE}$ from the complex analysis for the first three
P$_1$-branch lines (derived from different band-specific line sets) are shown
in Fig.~\ref{fig:delTnlte_vp}. The values with a conspicuous maximum of
$13.2 \pm 2.0$\,K at $v^{\prime} = 8$ are clearly higher than those from the
two-component fit. However, \citet{noll16} used HITRAN Einstein-A
coefficients, which significantly deviate from our modified \mbox{B+16}
coefficients. As demonstrated by Fig.~\ref{fig:Trot_Aset}, the impact on
$T_\mathrm{rot}$ can be large. Hence, we recalculated the
$\Delta T_\mathrm{NLTE}$ of \citet{noll16} with the modified \mbox{B+16}
transition probabilities for the lines considered in Sect.~\ref{sec:corrcoeff}
and the original ones (which result in about 2\,K higher $T_\mathrm{rot}$ on
average) in all other cases. Figure~\ref{fig:delTnlte_vp} indicates a clear
reduction of $\Delta T_\mathrm{NLTE}$ for the relevant $v^{\prime} \ge 4$, which
better matches our results based on two-component population fits. Between
$v^{\prime}$ of 4 and 7 the non-LTE contributions are almost constant with a
mean of 2.5\,K. However, the absolute uncertainties are larger. Only the
maximum of $8.4 \pm 2.7$\,K at $v^{\prime} = 8$ seems to be significant. It
might also be present (but less pronounced) in the population fitting results. 
The high absolute uncertainties from temperature comparisons
($v^{\prime}$-related differences are safer) are a critical drawback of that
method and imply that two-component population fits provide the best
constraints for $\Delta T_\mathrm{NLTE}$, so far. The higher errors compared
to the original \citet{noll16} data are partly related to the unavoidable
mixture of corrected and uncorrected \mbox{B+16} coefficients. However,
\mbox{B+16} line parameters also appear to cause a larger scatter in
$T_\mathrm{rot}$ for different bands with the same $v^{\prime}$ compared to
HITRAN data. The change in the $\Delta T_\mathrm{NLTE}$ differences between
adjacent $v^{\prime}$ (especially around $v^{\prime} = 7$) is mainly caused
by a different calculation of $T_\mathrm{rot}$ for the reference line set
consisting of the first three P$_1$-branch lines. Instead of using a constant
temperature offset for the conversion from the reference line set of
\citet{noll15} including all P-branch lines up to $N^{\prime} = 3$ as
discussed by \citet{noll16}, we directly corrected the band-specific
$T_\mathrm{rot}$ to be representative of the more recent reference line set.

\subsection{Population variability}\label{sec:popvar}

The discussion of the roto-vibrational level populations of OH in
Sect.~\ref{sec:poptrot} was only based on line intensity measurements in a
single mean spectrum. We can learn more about these populations if we also
consider variations in the emission layer properties. In order to keep the
signal-to-noise ratios high, we split the sample into two parts based on a
characteristic layer parameter, calculated the corresponding mean spectra, and
derived level populations from the measured line intensities for a comparison.
For the split, we selected the effective height of the OH emission layer
$h_\mathrm{eff}$, i.e. the centroid altitude weighted by the volume emission
rate, as it is positively correlated with the strength of the non-LTE effects
\citep{noll17,noll18a}. There are fewer thermalising collisions without
$v^{\prime}$ change at higher altitudes due to lower air densities but higher
atomic oxygen mixing ratios \citep{noll18b}. The impact of this effect is
clearly reflected by the observed higher $h_\mathrm{eff}$ for higher
$v^{\prime}$ \citep[e.g.,][]{savigny12}, which are more affected by the hot
nascent population and have lower effective lifetimes. According to the
population modelling of \citet{noll18b} for $v^{\prime} = 9$, $h_\mathrm{eff}$
also increases for higher $N^{\prime}$.

In order to study the change of the population distribution for the different
$v^{\prime}$ depending on the OH emission altitude, we need adequate
space-based measurements of the emission profiles to be linked with our
ground-based OH level population data. This was already achieved by
\citet{noll17} based on limb-sounding data for the Cerro Paranal region
from the OH-specific channels of the SABER radiometer \citep{russell99}. Here,
we focus on the channel centred on 2.06\,$\mu$m, which covers \mbox{OH(8-6)}
and \mbox{OH(9-7)}. The effective $v^{\prime}$ is about 8.3 \citep{noll16}.
\citet{noll17} connected the resulting $h_\mathrm{eff}$ for 4,496 profiles to
each UVES spectrum by a weighting procedure which involved temporal
differences in day of year and local time measured in a two-dimensional
climatology. The approach also included the correction of differences in the
solar activity, as measured by the solar radio flux
(Sect.~\ref{sec:lineint}), for each UVES observation compared with the
corresponding weighted $h_\mathrm{eff}$. Finally, a most likely
$h_\mathrm{eff}$ was available for each UVES spectrum. We used these data to
split our sample of 533 spectra (Sect.~\ref{sec:data}) at a median
$h_\mathrm{eff}$ of 89.2\,km for the 2.06\,$\mu$m OH channel. The resulting
subsamples show mean $h_\mathrm{eff}$ of 88.7 and 89.6\,km, i.e. the height
difference is almost 1\,km. The situation is very similar for the other OH
channel at 1.64\,$\mu$m representing an effective $v^{\prime}$ of about 4.6
\citep{noll16}, where the corresponding heights are 87.3 and 88.3\,km. Then,
we performed the entire data analysis starting with the calculation of the
mean spectra up to the derivation of the final populations. In order to
minimise systematic effects in the line measurement, the same wavelengths for
the line integration and continuum derivation as for the full sample spectrum
were used (see also Sect.~\ref{sec:lineint}).  

\begin{figure}[t]
\includegraphics[width=8.3cm]{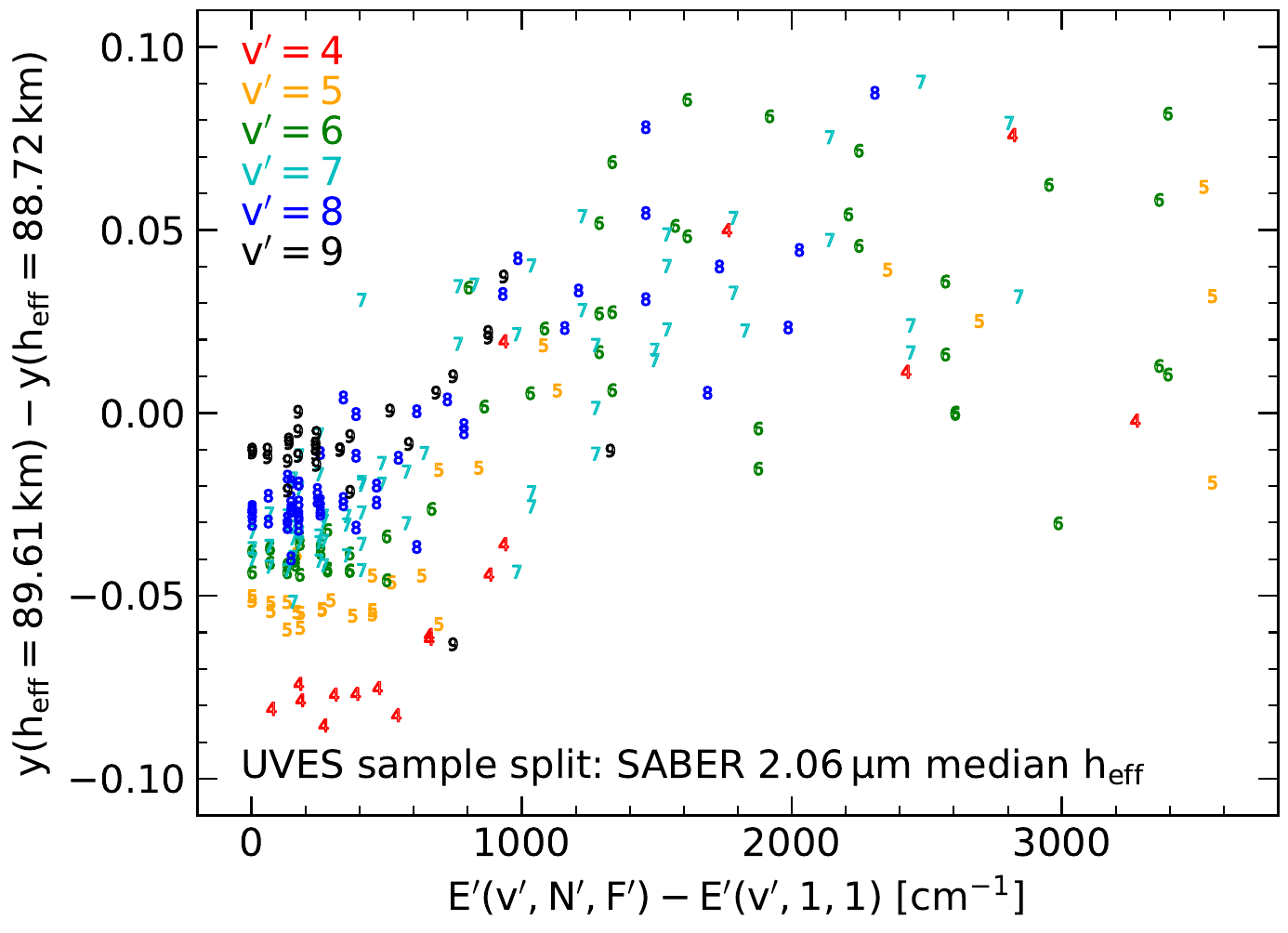}
\caption{Difference in logarithmic OH level populations for UVES mean spectra
  based on equally sized subsamples of high and low effective emission height
  (89.61 and 88.72\,km on average). The altitudes were derived from volume
  emission profiles of the SABER OH channel at 2.06\,$\mu$m for the Cerro
  Paranal region. The data points are shown for different upper vibrational
  levels $v^{\prime}$ (indicated by the corresponding coloured markers) as a
  function of the level energies in inverse centimetres relative to the lowest
  $v^{\prime}$-related energy. The plot only shows population measurements for
  reliable $\Lambda$ doublets which are covered by both UVES set-ups (see
  Fig.~\ref{fig:spec} for the valid wavelength ranges).}
\label{fig:dy_dEp_heff}
\end{figure}

Figure~\ref{fig:dy_dEp_heff} shows the resulting population ratios for the
high and low $h_\mathrm{eff}$ cases. For such a comparison, the choice of the
Einstein-A coefficients does not matter. We only plot $\Delta y$ related to
the 257 best-quality $\Lambda$ doublets (class 33) that are covered by all
UVES spectra, i.e. reliable population ratios which are representative of the
given $h_\mathrm{eff}$. We can identify three energy level regimes with respect
to the population change by a rise of the OH emission layer.

Up to about 600\,cm$^{-1}$, there is a general decrease of the populations,
which strongly depends on $v^{\prime}$. From $v^{\prime} = 4$ to 9, the decrease
shrinks from about 8\% ($-0.08$) to about 1\%, i.e. the populations for higher
$v^{\prime}$ are more stable. The largest difference in $\Delta y$ is between
the two lowest $v^{\prime}$. The decrease of the OH intensity for a rising OH
emission layer is well known \citep{yee97,melo99,liu06}. It is accompanied by
a lower width of the layer ($-0.5$\,km for our subsamples) due to especially
low OH production rates at the bottom side, which are caused by a depletion of
ozone. As band emissions with lower $v^{\prime}$ peak at lower altitudes
\citep{savigny12,noll16}, they seem to be more affected by this lack of fuel
for the OH production.

At $\Delta E^{\prime}$ above 1,200\,cm$^{-1}$, Fig.~\ref{fig:dy_dEp_heff} shows
a completely different behaviour. There is a general increase of the
populations with a mean of $+0.04$ and no significant dependence on
$v^{\prime}$. This finding implies that the contribution of hot populations to
the total population increases with $h_\mathrm{eff}$. The impact of non-LTE
effects grows by a less efficient thermalisation process. The reduced
contributions from lower altitudes to the total emission certainly play an
important role here since (similar to $v^{\prime}$) emission related to higher
$N^{\prime}$ peaks higher in the atmosphere \citep{noll18b}. There, collisional
thermalisation of the rotational level population distributions is hampered by
a relatively low density of nitrogen molecules and a relatively high volume
mixing ratio of $v^{\prime}$-deactivating (or even OH-destroying) atomic oxygen
radicals \citep{noll18b}. Note that the location of the zero line in
Fig.~\ref{fig:dy_dEp_heff} is uncertain with respect to the degree of
thermalisation as the increase of 4\% could also be caused by a change in
the OH column density. If the hot populations define the zero line as they
appear to be the most stable ones (which might be supported by the lack of a
$v^{\prime}$ dependence), the low-$N^{\prime}$ populations would further
decrease. 
 
Figure~\ref{fig:dy_dEp_heff} is another good argument for the bimodality of
rotational level population distributions. The change of $y$ for low and high
$N^{\prime}$ with increasing $\Delta E^{\prime}$ seems to be very small. This
suggests that cold and hot populations are relatively homogeneous, which
supports our two-temperature fit approach. Moreover, there is a quick
transition between both populations in the relatively narrow
$\Delta E^{\prime}$ range between 600 and 1,200\,cm$^{-1}$. In
Fig.~\ref{fig:ycorr_dEp}, this is the region where both fit components
significantly contribute. Hence, a rise of the OH emission layer there should
have the strongest impact on the slope of the population distribution (i.e.
$T_\mathrm{rot}$) by a change of the relative contribution of the cold and hot
populations. As $r_\mathrm{pop,0}$ increases, there is also an effect on
$\Delta T_\mathrm{NLTE}$ as shown in Fig.~\ref{fig:delTnlte_vp}. Estimates of
this quantity based on the populations for low and high OH layer are
relatively uncertain due to the distinctly lower number of suitable lines and
the high impact of increased line intensity errors on the analysis of very
small population differences. Nevertheless, we calculated
$\Delta T_\mathrm{NLTE}$ changes of the order of a few tenths of a kelvin for
the altitude difference of about 1\,km. This is clearly smaller than about
1\,K per kilometre, the order of magnitude from observational and modelling
studies by \citet{noll17,noll18a,noll18b}, which might point to limitations in
the study of $\Delta T_\mathrm{NLTE}$ variations based on two-component
population fits.

\conclusions[Conclusions]  
\label{sec:conclusions}

Based on averaged high-quality high-resolution spectra from the UVES echelle
spectrograph at Cerro Paranal, we performed a detailed study of OH
roto-vibrational level population distributions. The mean populations for 723
$\Lambda$ doublets with upper vibrational levels $v^{\prime}$ between 3 and 9
and upper rotational levels $N^{\prime}$ up to 24 were investigated. In about
half the cases, the doublet components were measured separately. The line
wavelengths from literature \citep{rothman13,brooke16} turned out to be
sufficiently accurate in most cases. Only a small number of lines with high
$N^{\prime}$ and intermediate $v^{\prime}$ (especially $v^{\prime} = 5$) showed
deviations by more than a few picometres.

The quality of population measurements is limited by uncertainties in the
Einstein-A coefficients. We investigated this issue with comparisons of
populations from different transitions with the same upper state. We tested
six sets of transition probabilities: \citet{brooke16}, HITRAN
\citep{rothman13}, \citet{vanderloo08}, \citet{turnbull89},
\citet{langhoff86}, and \citet{mies74}. All sets fail in the case of Q-branch
lines and the $\Lambda$-doublet components, where unexpectedly large intensity
ratios are possible. The comparison of populations from P- and R-branch lines
indicated relatively small errors for the coefficients by \citet{langhoff86},
\citet{vanderloo08}, and \citet{brooke16}, whereas those from
\citet{turnbull89} are clearly the worst. The comparison of OH bands with the
same $v^{\prime}$ showed a similar order of the different sets with respect to
their quality. For this case, the coefficients of \citet{brooke16} performed
best. The widely used HITRAN data are only of intermediate and hence
unsatisfactory quality.

For the population analysis, we focused on the Einstein-A coefficients from
\citet{brooke16} due to their relatively good performance and the highest
number of included lines. In order to minimise the scatter in the populations,
we further improved these coefficients by empirically correcting the found
population discrepancies via regression lines related to $N^{\prime}$ and
wavelength as well as band-dependent correction factors. For the correction of
the branch-related differences, we used P- and R-branch data combined with
equal weights as the reference since this strongly reduced the deviations
between the different sets of Einstein-A coefficients with respect to
rotational temperatures $T_\mathrm{rot}$, i.e. the change of the populations
with increasing $N^{\prime}$. The whole correction procedure lowered the
discrepancies in the coefficients by more than a factor of 2 for the measured
lines. Nevertheless, the development of an improved set for all lines would
need a more sophisticated approach including modelling of the molecular
parameters.

The resulting $v^{\prime}$-dependent population distributions show clearly
bimodal structures, which were convincingly reproduced by two-temperature fits
only excluding steep population decreases for $v^{\prime} = 8$ and 9 at the
highest $N^{\prime}$ with energies slightly below and above the exothermicity
limit of the OH-producing hydrogen--ozone reaction, respectively. The fits
show a cold population with nearly ambient temperature of about 190\,K
dominating at low $N^{\prime}$ and a hot population with temperatures between
700\,K for $v^{\prime} = 9$ and 7,000\,K for $v^{\prime} = 4$ at high
$N^{\prime}$. In contrast, the ratio of the hot and cold populations at the
level with the lowest energy of a given $v^{\prime}$ changes from 3 to 0.3\%
mainly due to a decrease of the cold component. The significant contribution
of a hot population to low $N^{\prime}$ causes deviations between
$T_\mathrm{rot}$ and ambient temperature, which we estimated by fitting our
two-component model for the energy levels related to the first three
P$_1$-branch lines. The results indicate non-LTE contributions that increase
from about 1\,K for $v^{\prime} = 4$ to about 5\,K for $v^{\prime} = 8$. The
best-fit value for $v^{\prime} = 9$ is even higher (about 6\,K), but the fit
uncertainties are by far the highest. In general, the applied approach is much
more robust than the previously used method based on comparisons of
temperatures from different sources as it only weakly depends on uncertainties
in the line intensities and Einstein-A coefficients. Our approach is mostly
limited by the reliability of the assumption of only two Boltzmann-like
population distributions. There are hints of the existence of a more complex
pattern but the impact of these additional components appears to be small.

This conclusion is supported by the change of the populations due to a rise of
the OH emission layer, which we studied by the separation of the sample of
spectra into two parts depending on the effective emission height as obtained
from height-resolved SABER OH volume emission rates. The energy regimes up to
about 600\,cm$^{-1}$ and above about 1,200\,cm$^{-1}$ relative to the lowest
energy for a given $v^{\prime}$ show clearly distinct variability in agreement
with the energy ranges dominated by the cold and hot components in the derived
population distributions. While the cold populations show a decrease, which is
stronger for lower $N^{\prime}$, the hot populations are relatively stable (or
even increase) with increasing emission altitude. The largest measured effect
is a 12\% decrease of the cold population at $v^{\prime} = 4$ relative to the
hot population for a height difference of almost 1\,km.

The success of the two-component model for OH rotational level population
distributions has implications for the thermalisation process of the highly
non-thermal nascent populations. There are still high uncertainties with
respect to the rate coefficients for collisions with and without change of
$v^{\prime}$. In particular, the modification of the rotational level
population by $v^{\prime}$-changing collisions is not known. Hence, the origin
of the very hot populations at high $N^{\prime}$ of low $v^{\prime}$ is puzzling.
Consequently, there is hope that the high-quality population data of this
study can help to better understand relaxation processes in OH by detailed
modelling. This will be important knowledge with respect to the use of OH as
an indicator of mesopause temperatures and for retrievals of atomic abundances
like those of oxygen.



\dataavailability{This project made use of the ESO Science Archive Facility at
\mbox{http://archive.eso.org}. UVES Phase\,3 spectra from different observing
programmes of the period from April 2000 to March 2015 were analysed. The v2.0
SABER data products used for this study were taken from
\mbox{http://saber.gats-inc.com}.}













\authorcontribution{S.~Noll has developed the project, processed the data,
  performed the analysis, produced the figures, and is the main author of the
  paper text, where all co-authors have made significant contributions.
  H.~Winkler and O.~Goussev have also influenced the design of the study. In
  addition, H.~Winkler has checked parts of the analysis and B.~Proxauf has
  been involved in the post-processing of the UVES Phase~3 products.
}

\competinginterests{The authors declare that they have no conflict of
  interest.}


\begin{acknowledgements}
  We thank reviewer E.~Oliva and one anonymous referee for their positive and
  helpful reports.
  S.~Noll is financed by the project NO\,1328/1-1 of the German Research
  Foundation (DFG). H.~Winkler is funded by the DFG project NO 404/21-1.
\end{acknowledgements}







\bibliographystyle{copernicus}
\bibliography{Nolletal2020a.bib}

\end{document}